\documentclass[aps,floatfix,superscriptaddress,showpacs,showkeys]{revtex4}

\usepackage{amssymb,amsmath,amsfonts,latexsym,graphicx,epsfig,
  textcomp,fancybox,mdframed}

\usepackage{color}
\usepackage{titlesec}
\titleformat{\section}{\large\bfseries}{\thesection}{1em}{}

\newcommand{\bea}{\begin{eqnarray}}
\newcommand{\ena}{\end{eqnarray}}
\newcommand{\nn}{\nonumber\\}
\newcommand\pfrac[2]{\left(\frac{#1}{#2}\right)}
\newcommand{\be}{\begin{equation}}
\newcommand{\en}{\end{equation}}
\newcommand{\imag}{\mathop{\rm Im}\nolimits}
\newcommand{\real}{\mathop{\rm Re}\nolimits}
\newcommand{\MeV}{\mathop{\rm MeV}\nolimits}
\newcommand{\GeV}{\mathop{\rm GeV}\nolimits}
\newcommand{\permille}{\,\text{\textperthousand}}

\begin{document}

\hfill MITP/19-042 (Mainz)

\title{A new angle on an old problem: Helicity approach to 
neutron beta decay \\ in the Standard Model}

\author{Stefan Groote}
\affiliation{F\"u\"usika Instituut, Tartu \"Ulikool,
  W.~Ostwaldi 1, EE-50411 Tartu, Estonia}
\author{J\"{u}rgen G. K\"{o}rner}
\affiliation{PRISMA$^+$ Cluster of Excellence, Institut f\"ur Physik,\\
  Johannes Gutenberg-Universit\"at, D-55099 Mainz, Germany}
\author{Bla\v zenka Meli\'c}
\affiliation{
Institut Rudjer Bo\v skovi\'c, Division of Theoretical Physics,
 Bijeni\v cka 54, HR-10000 Zagreb, Croatia.}

\today

\begin{abstract}
We elaborate on the dichotomy between the description of the semileptonic
decays of heavy hadrons on the one hand and the semileptonic decays of light
hadrons such as neutron $\beta$ decays on the other hand. For example, almost
without exception the semileptonic decays of heavy baryons are described in
cascade fashion as a sequence of two two-body decays
$B_1 \to B_2 + W_{\rm off-shell}$ and $W_{\rm off-shell} \to \ell + \nu_\ell$
whereas neutron $\beta$ decays are analyzed as true three-body decays
$n \to p + e^- +\bar \nu_e$. Within the cascade approach it is possible to
define a set of seven angular observables for polarized neutron $\beta$ decays
as well as the longitudinal, transverse and normal polarization of the decay
electron. We determine the dependence of the observables on the usual vector
and axial vector form factors. In order to be able to assess the importance of
recoil corrections we expand the rate and the $q^2$ averages of the
observables up to NLO and NNLO in the recoil parameter
$\delta=(M_n-M_p)/(M_n+M_p)= 0.689\cdot 10^{-3}$. Remarkably, we find that the
rate and three of the four parity conserving polarization observables that
we analyze are protected from NLO recoil corrections when the second class
current contributions are set to zero.
\end{abstract}

\maketitle

\section{Introduction\label{sec1}}
In the last few years there has been an extraordinary amount of activity on the
analysis of heavy baryon and heavy meson semileptonic decays 
$B_1(M_1) \to B_2(M_2) + \ell +\nu$ and rare decays
$B_1(M_1) \to B_2(M_2) + \ell^+ +\ell^-$  where $\ell=e,\mu,\tau$.
This ever-increasing activity has been fuelled by possible signals of lepton
flavour violation in these semileptonic or rare decay processes. Almost without
exception the analysis of the semileptonic and rare  heavy hadron decays was
done in cascade fashion where the decays were treated as a sequence of two
two-body decays~\cite{Frampton:1971sj,Korner:1987kd,Korner:1989ve,%
Korner:1989qb,Hagiwara:1989zt,Hagiwara:1989cu,Hagiwara:1989gza,Bialas:1992ny,%
Kadeer:2005aq,Faessler:2002ut,Feldmann:2011xf,Fajfer:2012vx,Gutsche:2013pp,%
Gutsche:2015mxa,Fischer:2018lme,Becirevic:2019tpx,Descotes-Genon:2019dbw,
Penalva:2019rgt,Ferrillo:2019owd,Das:2019omf,Cohen:2019zev,Mu:2019bin}. For
example, the semileptonic meson decay $M_1 \to M_2 + \ell + \nu$ is described
by the first stage two-body decay $M_1 \to M_2 + W_{\rm off-shell}$ followed
by the second stage two-body decay $W_{\rm off-shell} \to \ell + \nu$. In
contrast to this, neutron $\beta$ decays and semileptonic hyperon decays have
traditionally been analyzed in terms of the basic three-body decay process
$B_1 \to B_2 + \ell +\nu$~\cite{Lee:1956qn,Jackson:1957zz,Jackson:1957auh,%
Wilkinson:1982hu,Abele:2008zz,Nico:2009zua,Dubbers:2011ns,Vos:2015eba}. In
this paper we wish to demonstrate that there are many advantages in also
treating neutron $\beta$ decays as a cascade decay process
$n \to p + W^-_{\rm off-shell}$ followed by
$W^-_{\rm off-shell}\to e^- +\bar \nu_e$. In the first two-body decay
$n \to p + W^-_{\rm off-shell}$ the $W^-_{\rm off-shell}$ emerges polarized,
the polarization of which is subsequently analyzed by the second stage decay
$W^-_{\rm off-shell}\to e^- +\bar \nu_e$.

The advantage of the cascade approach to polarized neutron decays is that one
can define a larger number of unpolarized and neutron spin-related
polarization observables than is possible in the three-body decay approach.
One can count the number of independent hadronic helicity structure functions
that describe the quasi-two-body process
$n(\lambda_1) \to p(\lambda_2) +W^-_{\rm off-shell}(\lambda_W)$ by looking at
the independent elements of the hermitian double spin density matrix 
${\cal H}_{\lambda_W\, \lambda'_W}^{\,\lambda^{\phantom x}_1\,\lambda'_1}$. We
denote the helicities of the three particles involved in the quasi-two-body
decay by $\lambda_1, \lambda_2, \lambda_W$ such that
$\lambda_1 = \lambda_2 -\lambda_W$. One has to keep in mind that
$\lambda_1 +\lambda_W= \lambda_1' +\lambda_W'$ since one is not observing the
spin of the final state proton. Further one has
$|\lambda_1 +\lambda_W|= |\lambda_1' +\lambda_W'|= 1/2$ since the
helicity of the proton can only take the values $\lambda_2=\pm1/2$. The
helicity of the off-shell $W$ boson can assume the four values
$\lambda_W=t,+1,0,-1$ where $t$ denotes the time component of the off-shell
$W$ boson. There are thus altogether sixteen independent double spin density
matrix elements
\begin{eqnarray}
\label{doublespin}
&&{\cal H}_{--}^{++},\,{\cal H}_{++}^{--},\,{\cal H}_{00}^{++},\,%
{\cal H}_{00}^{--},\,\real {\cal H}_{-0}^{+-},\,\imag {\cal H}_{-0}^{+-},\,%
\real {\cal H}_{+0}^{-+},\,\imag {\cal H}_{+0}^{-+} \nn
&&{\cal H}_{tt}^{++},\,{\cal H}_{t0}^{++},\,{\cal H}_{tt}^{--},\,%
{\cal H}_{t0}^{--},\,\real {\cal H}_{-t}^{+-},\,\imag {\cal H}_{-t}^{+-},\,%
\real {\cal H}_{+t}^{-+},\,\imag {\cal H}_{+t}^{-+}.
\end{eqnarray}
We employ a concise notation for the double spin density matrix elements in
that we write $(\pm=\pm 1/2)$ for the upper indices $(\lambda_1\,\lambda_1')$
and $(\pm=\pm 1)$ for the lower indices $(\lambda_W\,\lambda_W')$. The set of
sixteen double spin density matrix elements in Eq.~(\ref{doublespin}) contains
twelve $T$-even and four $T$-odd structure functions. When counting the number
of polarization observables one has to subtract the trace of the double density
matrix since polarization observables correspond to normalized double spin
density matrix elements. This leaves one with eleven $T$-even and four $T$-odd
observables. In this paper we discuss a subset of seven angular and three
electron spin observables which are contributed to by linear combinations of
the above set of double spin density matrix elements. We are not exhausting
the full set of possible spin measurements which explains why the number of
our observables is smaller than the number of double spin density matrix
elements. For example, we do not consider the polarization of the final state
proton which would be very difficult to measure.

Compare this to the four independent single spin density matrix elements
\begin{equation}
\label{threebody1}
{\cal H}^{++},\,{\cal H}^{--},\, \real {\cal H}^{+-},\,\imag {\cal H}^{+-}
\end{equation}
of the polarized decay $n(\uparrow) \to p + e^- +\bar \nu_e$ where there are
three $T$-even and one $T$-odd structure functions. The relevant angular decay
distribution reads (see e.g.\ Ref.~\cite{Fischer:2018lme})
\begin{equation}
\label{threebody2}
\frac{d\Gamma_{\rm tot}}{d\cos\theta_P d\chi}\sim A+B\,P_n\cos\theta_P 
  +C\,P_n\sin\theta_P \cos\chi + D\,P_n\sin\theta_P \sin\chi
\end{equation}
where $A\sim{\cal H}^{++}+{\cal H}^{--}$, $B\sim {\cal H}^{++}-{\cal H}^{--}$,
$C\sim\real {\cal H}^{+-}$ and $D\sim\imag {\cal H}^{+-}$, and $P_n=|\vec P_n|$
is the magnitude of the polarization of the neutron. This leaves one with
the three independent normalized observables given by $B/A$, $C/A$ and $D/A$
compared to the 15 observables in the helicity approach. The angles $\theta_P$
and $\chi$ describe the orientation of the polarization vector $\vec P_n$ of
the neutron relative to the decay plane formed by the three final state
particles $(p,\,e^-,\,\bar \nu)$. The correlation angles and thereby the
correlation coefficients $B,\,C,\,D$ depend on the choice of the $z$ axis to
be in the decay plane or perpendicular to the plane (see e.g.\
Ref.~\cite{Fischer:2018lme}).

It must be clear that the physical content of the cascade approach and the
direct decay approach are the same but the physics appears in different guises
in the two approaches. By applying appropiate boosts, one can always convert
the results of one approach into the results of the other approach either
analytically or with the help of a Monte Carlo event generation program as
e.g.\ described in Ref.~\cite{Kadeer:2005aq}.

\section{Helicity and invariant amplitudes\label{sec2}}
We define the usual set of three parity conserving (p.c.) and three
parity violating (p.v.)
invariant form factors for the current-induced transition $n \to p$. One has
($\sigma_{\mu \nu}=i/2(\gamma_\mu\gamma_\nu-\gamma_\nu\gamma_\mu))$
\begin{eqnarray}
\label{covariant1}
M_\mu^V&=&\langle B_2|J_\mu^V|B_1\rangle
  \ =\ \bar{u}_p(p_2)\bigg[F_1^{V}(q^2)\gamma_\mu
  -i\frac{F_2^V(q^2)}{M_n}\sigma_{\mu \nu}q^\nu
  +\frac{F_3^V(q^2)}{M_n}q_\mu\bigg]u_n(p_1),\nonumber\\
M_\mu^A&=&\langle B_2|J_\mu^A|B_1\rangle
  \ =\ \bar{u}_p(p_2)\bigg[F_1^A(q^2)\gamma_\mu
  -i\frac{F_2^A(q^2)}{M_n}\sigma_{\mu \nu}q^\nu
  +\frac{F_3^A(q^2)}{M_n}q_\mu\bigg]\gamma_5 u_n(p_1),
\end{eqnarray}
where, differing from Ref.~\cite{Kadeer:2005aq}, we use the conventions of
Bjorken-Drell for the $\gamma$ matrices. In particular, we use
\begin{equation}
 \gamma_5=\left(\begin{array}{cc}0 & 1 \\ 1 & 0
\end{array}
\right).
\end{equation}
Next we linearly relate the six invariant form factors to six helicity
amplitudes $H^{V/A}_{\lambda_2\, \lambda_W}$ for the quasi-two-body process
$n(p_1,\lambda_1) \to p(p_2,\lambda_2) +W^-_{\rm off-shell}(q;\lambda_W)$
where $\lambda_1=\lambda_2-\lambda_W$. One obtains (see
e.g.\ Ref.~\cite{Kadeer:2005aq}) 
\begin{eqnarray}
\label{helamp}
H_{\frac{1}{2}1}^{V/A} &=&\sqrt{2Q_\mp}\bigg(
- F_1^{V/A}(q^2) \mp \frac{ M_\pm}{M_n} F_2^{V/A}(q^2) \bigg), \nn
H_{\frac{1}{2}0}^{V/A} &=& \frac{\sqrt{Q_\mp}}{\sqrt{q^2}}\bigg(
M_\pm F_1^{V/A}(q^2) \pm \frac{q^2}{M_n} F_2^{V/A}(q^2) \bigg), \nn
H_{\frac{1}{2}t}^{V/A} &=& \frac{\sqrt{Q_\pm}}{\sqrt{q^2}}\bigg(
M_\mp F_1^{V/A}(q^2) \pm  \frac{q^2}{M_n} F_3^{V/A}(q^2) \bigg).
\end{eqnarray}
We use the abbreviations $M_\pm = (M_n \pm M_p)$ and
$Q_\pm=(M_n \pm M_p)^2 - q^2$. The remaining helicity amplitudes are obtained
from the parity relations $H_{-\lambda_2, -\lambda_W}^{V/A} =
\pm H_{\lambda_2, \lambda_W}^{V/A}$.

At the zero-recoil point $q^2 = (M_n-M_p)^{2}$ only the $s$-wave
transitions survive. These are conventionally called allowed Fermi and allowed 
Gamow--Teller transitions, respectively. The surviving helicity amplitudes are
\begin{eqnarray}
H^{V}_{\frac12 t}&=&H^{V}_{-\frac12 t}
  \ =\ 2\sqrt{M_{1}M_{2}}\,(F_{1}^{V}+\frac{M_{-}}{M_{1}}F_{3}^{V})
  \qquad \qquad \qquad \qquad \qquad \qquad \qquad \mbox{allowed Fermi,}
\label{eq:fermi}\\
H^{A}_{\frac12 1}/\sqrt{2}&=&-H^{A}_{-\frac12 -1}/\sqrt{2}
  \ =\ H^{A}_{\frac12 0}=-H^{A}_{-\frac12 0}=2\,\sqrt{M_{1}M_{2}}\,(F_{1}^{A}
  -\frac{M_{-}}{M_{1}}F_{2}^{A})\qquad\mbox{allowed Gamow--Teller.}
\label{eq:gamow}
\end{eqnarray}
When one converts the helicity amplitudes to $(LS)$ amplitudes, one can see
that the above recoil relations project onto the $(LS)$ amplitudes
$(L=0, S=1/2)$ in both cases.

The ultimate goal in neutron $\beta$-decay experiments would be to measure
the complete set of six form factors $F_i^{V/A}\,(i=1,2,3)$ independent of
any theoretical input and then to confront the measurements with theoretical
expectations. This goal is difficult to realize because the contributions
of some of the form factors to the rate and to the polarization observables
are quite small and difficult to measure. In practise one concentrates
on the measurement of the axial form factor $F_1^A$ and, in that order,
on the weak magnetism form factor $F_2^V$.

Let us briefly list the theoretical expectations for five of the six form
factors that are based on i) the conserved vector current (CVC) hypothesis
determining the vector form factor $F_1^V(0)$ and the weak magnetism form
factor $F_2^V(0)$, ii) partial conservation of the axial vector current (PCAC)
specifying the induced pseudoscalar scalar form factor $F_3^A(0)$, and iii)
the absence of second class currents leading to the vanishing of the induced
scalar form factor $F_3^V(0)$ and the tensor form factor $F_2^A(0)$ as e.g.\
described in Ref.~\cite{Marshak1969}. One has
\begin{eqnarray}
\label{marshak}
F_1^V(0)&=&1, \qquad \qquad
F_2^V(0)=\tfrac 12 (\kappa_p -\kappa_n)=1.853, \qquad
F_3^V(0)=0 ,\nn
&&\qquad \quad \qquad F_2^A(0)=0, \qquad 
F_3^A(0)\,=\,2\, \frac{M_n^2}{m_\pi^2} F_1^A(0)\,=\,118.02,
  \quad\mbox{\cite{Marshak1969}}\nn
&&\hspace{4.2cm}F_3^A(0)\, = 175, \quad\mbox{\cite{Gonzalez-Alonso:2013ura}}
\end{eqnarray}
where we have included a second theoretical estimate of the induced scalar
form factor using some lattice data given in
Ref.~\cite{Gonzalez-Alonso:2013ura}.

The value of the axial form factor $F_1^A(0)$ is not determined by any general
theoretical argument. The PDG presents the measured value of $F_1^A(0)$ in
terms of the ratio
$\lambda=F_1^A(0)/F_1^V(0) =1.2724(23)$~\cite{Tanabashi:2018oca}. However,
since the CVC value of $F_1^V(0)=1$ is protected from first order symmetry
breaking by the Ademollo--Gatto theorem~\cite{Ademollo:1964sr}, we shall use
the PDG value for $F_1^A(0)/F_1^V(0)$ directly for $F_1^A(0)$. In our
numerical analysis we thus take the PDG-based value
\begin{equation}
\label{F1A}
F_1^A(0)=1.2724(23).
\end{equation}

The errors of the lattice calculations of the values of $F_1^A(0)$ have been
considerably reduced over the last few years and have reached the
$1\,\%$ level. Berkowitz et al.\ quote
$F_1^A(0)=1.278(21)(26)$~\cite{Berkowitz:2017gql} or from later papers by
the same lattice collaboration
$F_1^A(0)=1.271(13)$~\cite{Chang:2017oll,Chang:2018uxx}. Ottnad et al.\
quote $F_1^A(0)=1.251(24)$~\cite{Ottnad:2018fri}. The present situation
concerning lattice calculations of the neutron $\beta$-decay form factor
values is nicely summarized in Ref.~\cite{Aoki:2019cca}. In a non-lattice
calculation the authors of Ref.~\cite{Faessler:2008ix} have used a covariant
constituent quark model that incorporates chiral effects through a chiral
expansion to calculate $F_2^V(0)=1.853$, $F_1^A(0)=1.2695$ and
$F_3^A(0)=112.270$. In an explicit calculation the authors verified that the
model satisfies the Ademollo--Gatto theorem.

As we shall see in Sec.~\ref{sec4}, the $q^2$ dependence of the form factors
sets in only at NNLO or even at higher order in the recoil expansion. To the
accuracy we are aiming at one can therefore use $F_i^{V/A}(q^2)=F_i^{V/A}(0)$.
For the sake of brevity we shall always drop the argument in the form factors
and set $F_i^{V/A}$ for $F_i^{V/A}(0)$ everywhere.

The experimental measurement of the axial form factor $F_1^A$ is based on 
life-time measurements for which there are two differing results from either
beam measurements or from trap measurements of ultracold neutrons which differ
from each other by $4\,\sigma$ (see e.g.\ Ref.~\cite{Czarnecki:2018okw}). In
addition, there is a recent claim that the size of the radiative corrections
to neutron $\beta$ decay needed in the evaluation of $F_1^A$ was
underestimated in previous analysis'~\cite{Seng:2018yzq,Seng:2018qru} calling
into question the previously determined values of $F_1^A$. The present
situation thus calls for an independent measurement of $F_1^A$ which, in
addition, does not depend on the value of $V_{ud}$. In Sec.~\ref{sec6} we
therefore analyze the sensitivity of our set of observables to variations in
the input value of $F_1^A$. 

\begin{table}[ht] 
\begin{center}
\caption{Definition of helicity structure functions and their parity
  properties}
\def\arraystretch{2}
\begin{tabular}{ll}
\hline
parity-conserving (p.c.) & \qquad parity-violating (p.v.) \\
\hline
${\cal H}_U = {\cal H}_{F_-} = |H_{+\frac12 +1}|^2 + |H_{-\frac12 -1}|^2$
  \qquad & \qquad
${\cal H}_F = {\cal H}_{U_-} = |H_{+\frac12 +1}|^2 - |H_{-\frac12 -1}|^2$ \\
${\cal H}_L = |H_{+\frac12\, 0}|^2 + |H_{-\frac12\, 0}|^2$ \qquad & \qquad
${\cal H}_{L_-} = |H_{+\frac12\, 0}|^2 - |H_{-\frac12\, 0}|^2$ \\
${\cal H}_S = |H_{+\frac12\, t}|^2 + |H_{-\frac12\, t}|^2$ \qquad & \qquad
${\cal H}_{S_-} = |H_{+\frac12\, t}|^2 - |H_{-\frac12\, t}|^2$ \\
${\cal H}_{SL_-} = \real\left( H_{+\frac12\,0} H_{+\frac12\,t}^\dagger 
  - H_{-\frac12\,0} H_{-\frac12\,t}^\dagger \right) $ \qquad & \qquad
${\cal H}_{SL_+} = \real\left( H_{+\frac12\,0} H_{+\frac12\,t}^\dagger 
  + H_{-\frac12\,0} H_{-\frac12\,t}^\dagger \right) $ \\
\qquad & \qquad
  $ {\cal H}_{ISL_+} = \imag\left( H_{+\frac12\,0} H_{+\frac12\,t}^\dagger
  + H_{-\frac12\,0} H_{-\frac12\,t}^\dagger \right) $  \\
${\cal H}_{LT_+} = \real\left( H_{+\frac12\,+1}  H_{+\frac12\,0}^\dagger 
  + H_{-\frac12\,-1} H_{-\frac12\,0}^\dagger \right)$ \qquad & \qquad
${\cal H}_{LT_-} =  \real\left( H_{+\frac12\,+1} H_{+\frac12\,0}^\dagger 
  - H_{-\frac12\,-1} H_{-\frac12\,0}^\dagger \right)$ \\
${\cal H}_{ILT_+} = \imag\left( H_{+\frac12\,+1} H_{+\frac12\,0}^\dagger 
  + H_{-\frac12\,-1} H_{-\frac12\,0}^\dagger \right)$ \qquad & \qquad
${\cal H}_{ILT_-} =  \imag\left( H_{+\frac12\,+1} H_{+\frac12\,0}^\dagger 
  - H_{-\frac12\,-1} H_{-\frac12\,0}^\dagger \right)$ \\
${\cal H}_{ST_-} =  \real\left( H_{+\frac12\,+1} H_{+\frac12\,t}^\dagger 
- H_{-\frac12\,-1} H_{-\frac12\,t}^\dagger \right)$ \qquad & \qquad
${\cal H}_{IST_+} = \imag\left( H_{+\frac12\,+1} H_{+\frac12\,t}^\dagger 
  + H_{-\frac12\,-1} H_{-\frac12\,t}^\dagger \right)$  \\
${\cal H}_{\rm tot} = {\cal H}_U+{\cal H}_L +\delta_e \left( 3{\cal H}_S
  + {\cal H}_U + {\cal H}_L\right)$ \qquad & \qquad \\[2ex]
\hline
\end{tabular}
\label{tab:bilinears}
\end{center}
\end{table}
In Table~\ref{tab:bilinears} we list the bilinear forms of the helicity
amplitudes that appear in the four-fold angular decay distribution to be
discussed in the next section. The helicity amplitudes
$H_{\lambda_2\, \lambda_W}$ appearing in Table~\ref{tab:bilinears} refer to
the linear combination of the vector and axial vector helicity amplitudes
given by
\begin{equation}
H_{\lambda_2\,\lambda_W}
=H^V_{\lambda_2\,\lambda_W}-H^A_{\lambda_2\,\lambda_W}.
\end{equation}
The parity properties of the helicity structure functions given in
Table~\ref{tab:bilinears} follow from the parity transformation properties of
the vector current $V^\mu \sim (0^+,1^-)$ and the axial vector current
$A^\mu \sim (0^-,1^+)$ expressed as $J^P$ separately for the time and space
components. For the diagonal spin 0 - spin 0 and spin 1 - spin 1 contributions
the parity can be seen to be positive/negative for the sums/differences of
helicity bilinears with helicity labels $\lambda_2\,\lambda_W$ and
$-\lambda_2\,-\lambda_W$. For the nondiagonal spin 1 - spin 0 contributions
there is an extra minus sign resulting from the parity properties of the
currents.

The zero-recoil structure of the helicity amplitudes Eqs.~(\ref{eq:fermi})
and~(\ref{eq:gamow}) has implications for the helicity structure functions
listed in Table~\ref{tab:bilinears}. The eight p.v.\ helicity structure
functions vanish at zero recoil, i.e.\
\begin{equation}
\label{pvzero}
{\cal H}_{ILT_-},\,{\cal H}_{IST_+},\,{\cal H}_{SL_+},\,{\cal H}_{F},\,
{\cal H}_{L_-},\,{\cal H}_{S_-},{\cal H}_{LT_{-}},{\cal H}_{ISL_{+}}=0.
\end{equation}
Six of the seven p.c.\ helicity structure functions take the recoil values
\begin{eqnarray}
{\cal H}_{U}&=&2\,{\cal H}_{L}\ =\ \sqrt{2}\,{\cal H}_{LT_+}
  \ =\ 4M_nM_p\, |F_1^A|^2\,, \nn
{\cal H}_{ST_-}&=&2\,{\cal H}_{SL_-}
  \ =\ -8\sqrt{2}M_nM_p \,\real\left(F_1^V\,F_1^{A \ast}\right) \nn
{\cal H}_{S}&=& 8M_nM_p\,|F_1^V|^2, \qquad  {\cal H}_{ILT_+}=0\,,
\label{eq:zerec}
\end{eqnarray}
while the p.c.\ helicity structure function ${\cal H}_{ILT_+}$ is zero. As in
Eqs.~(\ref{eq:fermi}) and~(\ref{eq:gamow}) we have set the second class form
factors $F_3^V$ and $F_2^A$ to zero. One can check that the p.v.\ helicity
structure functions are proportional to $p$ when $F_3^V=F_2^A=0$ in agreement
with Eq.~(\ref{pvzero}). The zero-recoil relations can be used to quickly
assess the limiting behavior of the polarization observables in the
zero-recoil limit.

\section{Four-fold angular decay distribution\label{sec3}}
\begin{figure}
  \epsfig{figure=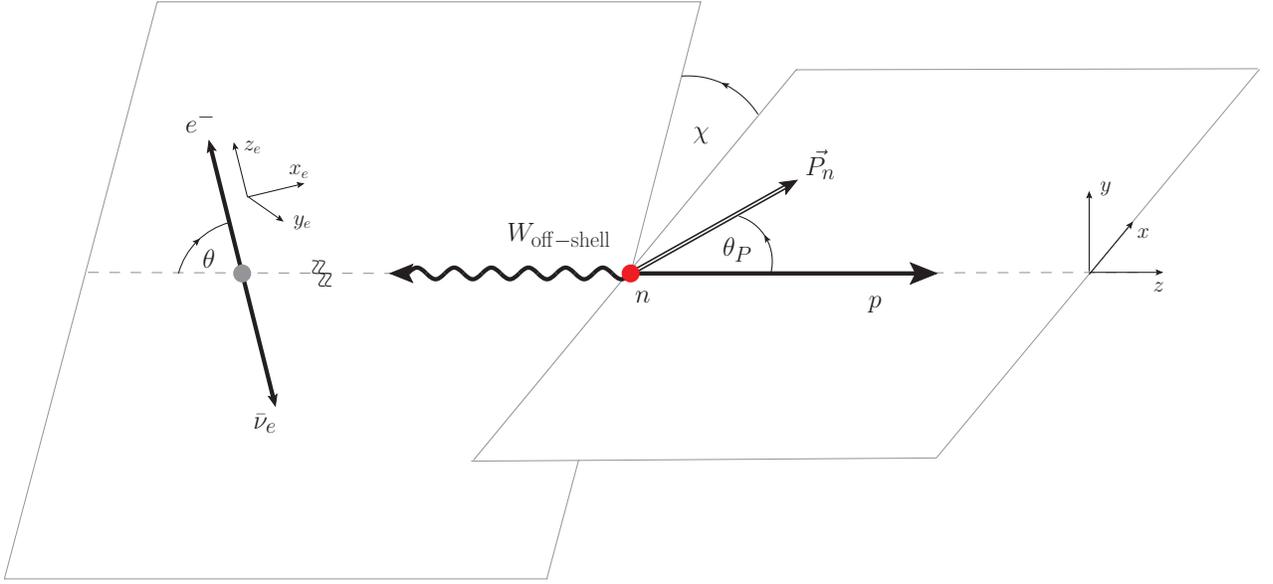, scale=0.40}
  \vspace{0.5cm}
\caption{\label{xyzsystem} Definition of the polar angles $\theta$ and
  $\theta_P$, and the azimuthal angle $\chi$ describing the decay of a
  polarized neutron using the lepton pair as polarization analyzer.
  $\vec{P_n}$ denotes the polarization vector of the neutron which is chosen
  to lie in the $(x,z)$ plane. The  components of the polarization vector of
  the electron $\vec P_e$ are defined in the right-handed $(x_e,y_e,z_e)$
  coordinate frame. The $y_e$--axis points into the page.}
\end{figure}
The angular decay distribution $W(\theta,\theta_P,\chi)$ is determined by the
master formula (see e.g.\ Ref.~\cite{Kadeer:2005aq})
\begin{eqnarray}
\label{master1}
W(\theta,\theta_P,\chi) &\propto&
  \sum_{\lambda_e,\lambda_W,\lambda'_W,J,J',\lambda_2} 
  \rho_{\lambda_2-\lambda_W,\lambda_2-\lambda'_W}(\theta_P)(-1)^{J+J'}
  |h_{\lambda_e\lambda_{\bar\nu}=1/2}|^2
  e^{-i(\lambda_W-\lambda'_W)\chi} \nn && \strut\times
  d^J_{\lambda_W,\,\lambda_e-\lambda_{\bar\nu}}(\theta)
  d^{J'}_{\lambda'_W,\,\lambda_e-\lambda_{\bar\nu}}(\theta)
  H_{\lambda_2\lambda_W}H^*_{\lambda_2\lambda'_W},
\end{eqnarray}
where $\lambda_{\bar \nu}=1/2$ for the massless antineutrino and
$\lambda_{e}=\pm1/2$. In the configuration $\lambda_{e}=+1/2$ the electron
helicity is flipped whereas the helicity is not flipped when
$\lambda_{e}=-1/2$. The penalty factor for flipping the helicity of the
electron is given by the ratio of the squares of the relevant leptonic flip
and nonflip helicity amplitudes
$|h_{\lambda_e=1/2\,\lambda_{\nu}=1/2}|^2/
\,|h_{\lambda_e =-1/2\,\,\lambda_{\nu}=1/2}|^2 = m_e^2/2q^2\,=:\,\delta_e$.
The polarization of the neutron is described by the normalized density matrix
\begin{equation}
\label{densitym}
\rho_{\lambda_{1}\lambda_{1}^{'}}(\theta_P)=\frac12\left(
  \begin{array}{cc}
    1+P_n \cos\theta_P & P_n \sin\theta_P\\
    P_n \sin\theta_P & 1-P_n \cos\theta_P
   \end{array}
\right),
\end{equation}
where $P_n$ denotes the magnitude of the polarization of the neutron.
For completeness we list the Wigner small $d^1(\theta)$ function appearing in
Eq.~(\ref{master1}) which is given by 
\begin{equation}
d^1_{mm'}(\theta)=\left(\begin{array}{ccc}
\frac12(1+\cos\theta)&-\frac1{\sqrt{2}}\sin\theta&\frac12(1-\cos\theta) \\
\frac1{\sqrt{2}}\sin\theta&\cos\theta&-\frac1{\sqrt{2}}\sin\theta \\
\frac12(1-\cos\theta)&\frac1{\sqrt{2}}\sin\theta&\frac12(1+\cos\theta) 
\end{array}\right)\vspace{0.1cm}
\end{equation}
The rows and columns are labeled in the order of $(1,0,-1)$. The spin-0 Wigner
function is simply $d^0_{00}(\theta)=1$.

The angles $\theta$, $\theta_P$ and $\chi$ are defined in Fig.~\ref{xyzsystem}.
Fig.~\ref{xyzsystem} provides a clear visualization of the two reference
frames used in the cascade analysis. For once, there is the neutron rest frame
which we will refer to as the $n$ frame, and second, one has the
$W_{\rm off-shell}$ rest frame (or $(e^-\,\bar \nu)$ center-of-mass frame)
which  will be referred to as the $q$ frame.
 
We mention that we have checked the correctness of the leptonic part of
the angular decay distribution~(\ref{master1}) given by
\begin{equation}
{\cal L}_{\lambda_W\,\lambda'_W}^{J\,J'}(\lambda_e)
  = |h _{\lambda_e\lambda_{\bar\nu}=1/2}|^2 e^{-i(\lambda_W-\lambda'_W)\chi}  
  d^J_{\lambda_W,\,\lambda_e-\lambda_{\bar\nu}}(\theta)
  d^{J'}_{\lambda'_W,\,\lambda_e-\lambda_{\bar\nu},}(\theta)
\end{equation}
by an independent covariant calculation.

Putting in the correct normalization one obtains the four-fold decay
distribution which we write as
\begin{equation}
\label{angdist1}
\frac{d\Gamma_{\rm tot}}{dq^2 d\cos\theta d\chi d\cos\theta_P}=
\frac{1}{4\pi}\frac{\Gamma_0(q^2-m_e^2)^2p}{M_n^7q^2}
\Big(W(\theta)+P_n\cdot W^P(\theta,\theta_P,\chi) \Big)\,,
\end{equation}
where
\begin{equation}
\Gamma_0 =\frac{G^2|V_{ud}|^2 M_n^5}{192\pi^3}.
\end{equation}
The momentum factor $p$ denotes the magnitude of the three-momentum of the
proton or of the $W^-_{\rm off-shell}$ boson given by $p=\sqrt{Q_+Q_-}/2M_n$.

The $\cos\theta$-dependent unpolarized decay distribution $W(\theta)$ and the
$(\theta,\theta_P,\chi)$-dependent three-fold polarized angular decay
distributions $W^P(\theta,\theta_P,\chi)$ can be calculated from
Eq.~(\ref{master1}). One has
\begin{eqnarray}
\label{angdist2}
W(\theta)&=&
  \frac{3}{8}(1+\cos^2\theta)(|H_{\frac{1}{2}1}|^2 +|H_{-\frac{1}{2}-1}|^2)
  -\frac{3}{4}\cos\theta (|H_{\frac{1}{2}1}|^2 -|H_{-\frac{1}{2}-1}|^2)  
  +\frac{3}{4}\sin^2\theta\Big(|H_{\frac{1}{2}0}|^2
  +|H_{-\frac{1}{2}0}|^2\Big) \nn &&
  +\frac{m_e^2}{2q^2}
  \bigg\{\frac{3}{2}(|H_{\frac{1}{2}t}|^2+|H_{-\frac{1}{2}t}|^2) 
  -3\cos\theta \, \real\left( H_{\frac{1}{2}t}H^\ast_{\frac{1}{2}0}
  + H_{-\frac{1}{2}t}H^\ast_{-\frac{1}{2}0}\right) \nn && \qquad
  +\frac{3}{2}\cos^2\theta (|H_{\frac{1}{2}0}|^2
  +|H_{-\frac{1}{2}0}|^2 )
  +\frac{3}{4}\sin^2\theta(|H_{\frac{1}{2}1}|^2
  +|H_{-\frac{1}{2}-1}|^2) \bigg\} 
\end{eqnarray}
and 
\begin{eqnarray}
\label{angdist3}
\lefteqn{W^P(\theta,\theta_P,\chi)\ =}\nn &&
\bigg[-\frac{3}{8}(1+\cos^2\theta)(|H_{\frac{1}{2}1}|^2-|H_{-\frac{1}{2}-1}|^2)
  +\frac{3}{4}\cos\theta (|H_{\frac{1}{2}\,1}|^2 +|H_{-\frac{1}{2}-1}|^2) 
  +\frac{3}{4}\sin^2\theta\Big(|H_{\frac{1}{2}0}|^2
  -|H_{-\frac{1}{2}0}|^2\Big)\bigg]\cos\theta_P \nn &&
  -\frac{3}{2\sqrt{2}} \sin\theta \sin\theta_P
  \bigg[\cos\chi \real\left(H_{\frac{1}{2}1} H^*_{\frac{1}{2}0}+
  H_{-\frac{1}{2}-1}H^*_{-\frac{1}{2}0}\right)+\sin\chi
  \imag \left( H_{\frac{1}{2}1} H^*_{\frac{1}{2}0}
  -H_{-\frac{1}{2}-1}H^*_{-\frac{1}{2}0}\right)\bigg] \nn &&
  +\frac{3}{4\sqrt{2}}\sin2\theta\sin\theta_P \bigg[\cos\chi
   \real\left( H_{\frac{1}{2}1}H^*_{\frac{1}{2}0}
  -H_{-\frac{1}{2}-1}H^*_{-\frac{1}{2}0}\right)
  +\sin\chi \imag \left( H_{\frac{1}{2}1}H^*_{\frac{1}{2}0}
  +H_{-\frac{1}{2}-1}H^*_{-\frac{1}{2}0}\right)\bigg] \nn &&
  +\frac{m_e^2}{2q^2}
  \bigg\{\frac{3}{2}\cos\theta_P(|H_{\frac{1}{2}t}|^2-|H_{-\frac{1}{2}t}|^2) 
  -3\cos\theta \cos\theta_P \real\Big( H_{\frac{1}{2}t}H^\ast_{\frac{1}{2}0}
  -H_{-\frac{1}{2}t}H^\ast_{-\frac{1}{2}0}\Big) \nn && \qquad
  +\frac{3}{2}\cos^2\theta \cos\theta_P
  \Big(|H_{\frac{1}{2}0}|^2-|H_{-\frac{1}{2}0}|^2\Big) 
  -\frac{3}{4}\sin^2\theta\cos\theta_P(|H_{\frac{1}{2}1}|^2
  -|H_{-\frac{1}{2}-1}|^2 ) \nn && \qquad
  +\frac{3}{\sqrt{2}} \sin\theta \sin\theta_P 
  \bigg[ \cos\chi\, \real(H_{\frac{1}{2}1}H^*_{\frac{1}{2}t}
  -H_{-\frac{1}{2}-1}H^*_{-\frac{1}{2}t})
  +\sin\chi\, \imag(H_{\frac{1}{2}1}H^*_{\frac{1}{2}t}
  +H_{-\frac{1}{2}-1}H^*_{-\frac{1}{2}t})\bigg] \nn && \qquad
  -\frac{3}{2\sqrt{2}}\sin 2 \theta \sin\theta_P 
  \bigg[\cos\chi\, \real (H_{\frac{1}{2}1}H^*_{\frac{1}{2}0}
  -H_{-\frac{1}{2}-1}H^*_{-\frac{1}{2}0})
  +\sin\chi\, \imag(H_{\frac{1}{2}1}H^*_{\frac{1}{2}0}
  +H_{-\frac{1}{2}-1}H^*_{-\frac{1}{2}0})\bigg] \bigg\},\qquad
\end{eqnarray}
where one has to remember to take the extra minus sign into account for the
spin 0 - spin 1 interference contributions in Eq.~(\ref{angdist3})
according to the factor $(-1)^{J+J'}$. This factor arises from having used the
Minkowski metric in the contraction of the lepton and hadron tensors (see
e.g.\ Ref.~\cite{Korner:1989ve}).
\begin{figure}
  \epsfig{figure=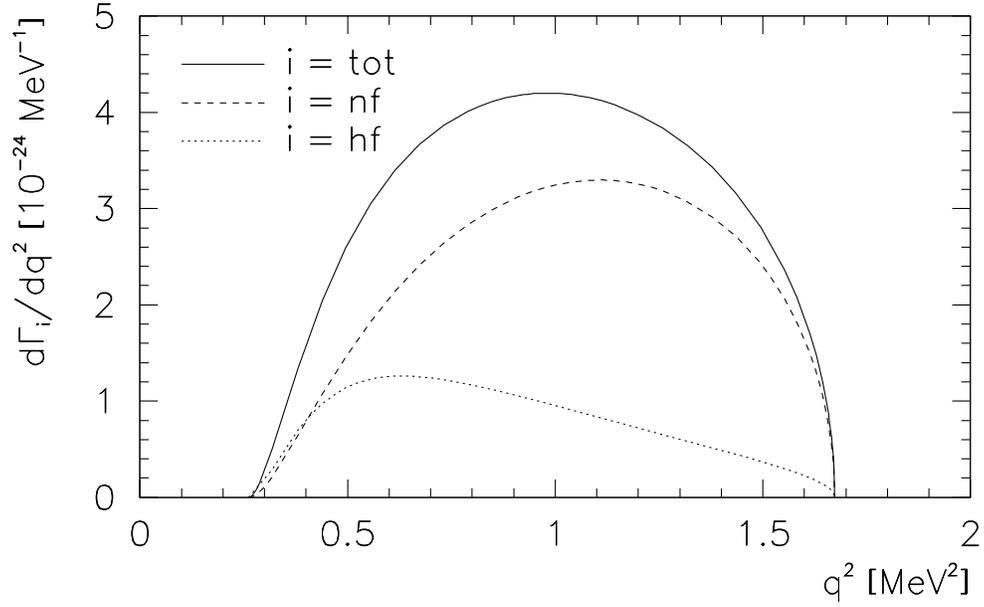,scale=0.8}
  \caption{\label{dgamma} $q^2$ dependence of the partial flip and non-flip
    rates and the total rate}
\end{figure}

Let us briefly pause to discuss some kinematical aspects of the problem. The
angle $\theta$ can be determined by measuring the energy $E_e$ of the electron
in the neutron rest frame from the relation (see Sec.~\ref{sec7})
\begin{equation}
\label{transtheta}
\cos\theta=\frac{1}{p(q^2-m_e^2)}\Big(2q^2E_e-q_0(q^2+m_e^2)\Big).
\end{equation}
This would require the knowledge of $q^2$, the value of which could be
determined from a measurement of the energy or momentum of the recoiling
proton. Barring hard photon emission from the neutron, proton or the
$W_{\rm off-shell}$, the relevant relations are $E_p=(M_n^2+M_p^2-q^2)/2M_n$
and $p=\sqrt{Q_+Q_-}/2M_n$. If the energy and/or the momentum of the
proton cannot be measured, one can determine $q^2$ by inverting the relation
$\cos\theta_{pe}=\cos\theta_{pe}(E_e,q^2)$ between the cosine of the opening
angle of the proton and the electron in the $n$ frame and the kinematic
variables $E_e$ and $q^2$ given by
\begin{equation}
\label{anglepe}
\cos\theta_{pe}=M_n\frac{2E_pE_e-M_n(E_p+E_e)+M_p^2+m_e^2}{\sqrt{Q_+Q_-}
  \,|\vec p_e|}.
\end{equation}
In this case one would have to take care to properly treat the two solutions
of the quadratic equation in $q^2$ that would result from inverting
Eq.~(\ref{anglepe}).

Returning to Eq.~(\ref{angdist3}), one notes that the angular factors
multiplying the imaginary parts of the helicity bilinears in
Eq.~(\ref{angdist3}) can be identified as $T$-odd triple momentum factors by
writing 
\begin{eqnarray}
\label{triple}
 \sin\chi\sin\theta\sin\theta_P &=&
  -\hat p_e \cdot (\hat s_n \times \hat p_p)  \nn
\sin\chi\sin2\theta\sin\theta_P
&=&  -2\,\,\hat p_e \cdot (\hat s_n \times \hat p_p)\,\hat s_n \cdot \hat p_p 
\end{eqnarray}
where $\hat s_n$ is a unit vector in the direction of the polarization of the
neutron and where the various unit three-vectors can be read-off from
Fig.~\ref{xyzsystem}. In explicit form they read $\hat p_p=(0,0,1)$,
$\hat s_n=(\sin\theta_P,0,\cos\theta_P)$ and
$\hat p_e=-(\sin\theta \cos\chi,\sin\theta \sin\chi,\cos\theta)$. It is not
difficult to see that the triple momentum product
$\hat p_e \cdot (\hat s_n \times \hat p_p)$ can be rewritten in the form
$ \hat p_e' \cdot (\hat s_n \times \hat p_{\hat \nu}' )$ where the primed
three-vectors refer to the corresponding three-vectors in the $n$ frame and
where one has used three-momentum conservation in the $n$ frame
$\hat p_p+\hat p_e'+\hat p_{\hat \nu}'=0$. The coefficient multiplying the
triple product $ \hat p_e' \cdot (\hat s_n \times \hat p_{\hat \nu}' )$ is
referred to as the $D$ term in the conventional three-body-decay approach. The
associated $T$-odd observables can be fed by true $CP$-violating contributions
or by $CP$-conserving electromagnetic rescattering corrections. In the
Standard Model the $CP$-violating contributions in the $d\to u$ sector are of
${\cal O}(10^{-12})$ and are thus negligibly small~\cite{Herczeg:1997se}. The
radiative rescattering corrections are also quite small. In the following we
therefore assume that the form factors $F_i^{V/A}$ are relatively real and
shall not further discuss the $T$-odd observables.

It is convenient and by now common practise to rewrite the angular decay
distribution~(\ref{angdist1}) in terms of the three Legendre polynomials
$P_0(\cos\theta)=1$, $P_1(\cos\theta)=\cos\theta$ and
$P_2(\cos\theta)=(3\cos^2\theta-1)/2$. One obtains
\begin{eqnarray}
  \label{angdist4}
  \frac{ d\Gamma_{\rm tot}}{dq^2 d\cos\theta d\cos\theta_P d\chi}
  &=&\frac{1}{8\pi}\Gamma_0 \frac{(q^2-m_e^2)^2p}{M_n^7q^2}
  \bigg\{{\cal H}_{\rm tot}(q^2) + {\cal H}_1(q^2) P_1(\cos\theta) +
  {\cal H}_2(q^2) P_2(\cos\theta) \nn &&
  + P_n \,  \cos\theta_P \Big ( {\cal H}_3(q^2) +
  {\cal H}_4(q^2) P_1(\cos\theta) + {\cal H}_5(q^2) P_2(\cos\theta) \Big)
  \nn &&
  + P_n \,  \cos\chi  \sin \theta_P \Big( \sin\theta  \, {\cal H}_6(q^2)
  + \sin 2\theta \, {\cal H}_7(q^2) \Big) \, \bigg\},
\end{eqnarray}
where
\begin{equation}
{\cal H}_{\rm tot}={\cal H}_U+{\cal H}_L
+\delta_e (3{\cal H}_S +{\cal H}_U+{\cal H}_L).
\end{equation}
The parity properties of the angular factors in Eq.~(\ref{angdist4}) are
determined by the parity transformations $\theta \to (\pi - \theta)$,
$\theta_P \to (\pi - \theta_P)$ and $\chi \to \chi + \pi$. The coefficients
${\cal H}_i(q^2)$ multiplying the angular factors are linear superpositions of
the helicity structure functions defined in Table~\ref{tab:bilinears}. For the
unpolarized case one has
\begin{eqnarray}
{\cal H}_1(q^2) &=& - \frac{3}{2} \Big({\cal H}_F
    + 4\, \delta_e {\cal H}_{SL_+}\Big), \qquad \quad \,\, \mbox{(p.v.)} \nn
{\cal H}_2(q^2) &=& \frac{1}{2} \left( 1 - 2\,\delta_e \right)
  \Big( {\cal H}_U - 2 {\cal H}_L \Big), \qquad \mbox{(p.c.)} 
\end{eqnarray}
and for the polarized case 
\begin{eqnarray}
{\cal H}_3(q^2) &=& -\left(1 + \,\delta_e\right)
  \left( {\cal H}_F - {\cal H}_{L_-} \right)
  + 3 \,\delta_e  {\cal H}_{S_-}, \qquad \mbox{(p.v.)} \nn
{\cal H}_4(q^2) &=& \frac{3}{2}
\left( {\cal H}_U - 4\,\delta_e {\cal H}_{SL_-} \right),
  \hspace{2.9cm} \mbox{(p.c.)} \nn
{\cal H}_5(q^2) &=& -\frac{1}{2} \left( 1 - 2\,\delta_e \right) 
  \left( {\cal H}_F + 2 {\cal H}_{L_-} \right),
  \hspace{1.7cm} \mbox{(p.v.)} \nn
{\cal H}_6 (q^2) &=& - \frac{3}{\sqrt{2}}
  \left( {\cal H}_{LT_+} - 2\,\delta_e  {\cal H}_{ST_-} \right),
  \hspace{2.0cm} \mbox{(p.c.)}\nn
{\cal H}_7(q^2)&=& \frac{3}{2 \sqrt{2}} \left( 1 - 2\,\delta_e  \right)
  {\cal H}_{LT_-}. \hspace{2.9cm} \mbox{(p.v.)}
\end{eqnarray}
The parity properties of the helicity structure functions indicated in round
brackets follow from the parity properties of the bilinear forms listed in
Table~\ref{tab:bilinears}.

Integrating the distribution~(\ref{angdist4}) over the three angles $\theta$,
$\theta_P$, and $\chi$ one obtains the total differential rate given by
\begin{equation}
\label{q2difftot}
\frac{d\Gamma_{\rm tot}}{dq^2}
  =\frac{\Gamma_0(q^2-m_e^2)^2p}{M_n^7q^2}\,{\cal H}_{\rm tot}(q^2).
\end{equation}
In analogy to Eq.~(\ref{q2difftot}) we define partial differential rates
according to
\begin{equation}
\label{q2partial}
\frac{d\Gamma_i}{dq^2}=\frac{\Gamma_0(q^2-m_e^2)^2p}{M_n^7q^2}\,
  {\cal H}_i(q^2).
\end{equation}
This leads to our final form of the angular decay distribution where we
factor out the total differential rate from the curly bracket in
Eq.~(\ref{angdist4}). One has
\begin{eqnarray}
\label{angdist5}
\frac{d\Gamma_{\rm tot}}{dq^2 d\cos\theta d\cos\theta_P d\chi}
  &=&\frac{1}{8\pi}\frac{d\Gamma_{\rm tot}}{dq^2}
  \bigg\{1 + {\cal O}_1(q^2) P_1(\cos\theta)
  + {\cal O}_2(q^2) P_2(\cos\theta) \nn &&
  + P_n \, \cos\theta_P \Big ( {\cal O}_3(q^2)
  + {\cal O}_4(q^2) P_1(\cos\theta) + {\cal O}_5(q^2) P_2(\cos\theta) \Big )
  \nn &&
  + P_n \, \cos\chi \sin \theta_P \Big( \sin\theta \, {\cal O}_6(q^2)
  + \sin 2\theta \, {\cal O}_7(q^2) \Big) \, \bigg\},
\end{eqnarray}
where the normalized observables ${\cal O}_i(q^2)$ are given by
\begin{equation}
{\cal O}_i(q^2)=\frac{{\cal H}_i(q^2)}{{\cal H}_{\rm tot}(q^2)}
   = \frac{d\Gamma_i\big/\,dq^2}{d\Gamma_{\rm tot} \big/\,dq^2}.
\end{equation}

Next we discuss how to isolate the individual angular observables from the
full decay distribution~(\ref{angdist4}). There are three principal ways to do
so. The most straightforward way is by a fit to the experimental angular decay
distribution. A second possibility is to project out the observables by taking
moments of the angular decay distribution w.r.t.\ appropiately chosen
trigonometric functions as in Refs.~\cite{Ivanov:2016qtw,Blake:2017une}. For
example, the $\cos\theta$ dependent terms can be projected out by folding with
Legendre polynomials. We follow a third method where one divides the angular
phase space into different sectors and takes piece-wise sums and differences
of the different sectors. This definition naturally leads to the set of
frequently discussed asymmetry parameters.

The first observable ${\cal O}_1(q^2)$ can be projected out by the standard
forward--backward projection
\begin{eqnarray}
\label{FB1}
A_{\rm FB}(q^2)\ =\ \frac 12 {\cal O}_1(q^2)
  &=&\frac{1}{d\Gamma_{\rm tot}/dq^2}\cdot\,\left (\int_{0}^{1}
  - \int_{-1}^{0} \right )d\cos\theta \int_{-1}^1 d\cos\theta_P
  \int_{0}^{2\pi}d\chi
  \frac{d\Gamma_{\rm tot}}{dq^2 d\cos\theta d\cos\theta_P d\chi}\nn
  &=& I_1(p.w.)\bigg[\left(\int_{0}^{1}-\int_{-1}^{0}\right)d\cos\theta\bigg],
\end{eqnarray}
where in the second row we have introduced a symbolic notation for the
piece-wise (p.w.) integration described in the first row. The symbolic
notation implies that the angular variables that do not appear in the
symbolic-integration symbol are integrated over their full range. Using the
symbolic notation one can project out the remaining observables in the angular
decay distribution~(\ref{angdist4}). One has
\begin{eqnarray}
\label{FB2}
A_{\rm conv}(q^2)\ = \frac 38 {\cal O}_2(q^2)
  &=&I_2(p.w.)\bigg[ \left(  \int_{1/2}^{1} - \int_{0}^{1/2}
  - \int_{-1/2}^{0}  + \int_{-1}^{-1/2} \right )d\cos\theta \bigg], \nn
A_{\rm PFB}(q^2)\ =\ P_n \cdot \frac 12 {\cal O}_3(q^2)
  &=&I_3(p.w.)\bigg[ \left(\int_{0}^{1}
  - \int_{-1}^{0} \right)d\cos\theta_P \bigg],\nn
A_{\rm DFB}(q^2)\ =\ P_n \cdot\frac 14{\cal O}_4(q^2)
  &=&I_4(p.w.)\bigg[\left(\int_{0}^{1}
  - \int_{-1}^{0} \right )d\cos\theta \bigg]
  \bigg[\left(\int_{0}^{1} - \int_{-1}^{0} \right)d\cos\theta_P \bigg], \nn    
A_{\rm Pconv}(q^2)\ =\ P_n \cdot\frac{3}{16}{\cal O}_5(q^2)
  &=&I_5(p.w.)\bigg[ \left( \int_{1/2}^{1} - \int_{0}^{1/2}
  - \int_{-1/2}^{0} + \int_{-1}^{-1/2} \right )d\cos\theta\bigg]
  \bigg[\left(\int_{0}^{1} - \int_{-1}^{0} \right)d\cos\theta_P \bigg], \nn
A_{\chi_1(q^2)}\ =\ P_n \cdot\frac{\pi}{8}{\cal O}_6(q^2)
  &=&I_6(p.w.)\bigg[2\left(\int_{0}^{\pi/2}
  - \int_{\pi/2}^{\pi} \right )d\chi\bigg], \nn
A_{\chi_2}(q^2)\ =\ P_n \cdot\frac{1}{3}{\cal O}_7(q^2)
  &=&I_7(p.w.)\bigg[ \left(\int_{0}^{1}
  - \int_{-1}^{0} \right )d\cos\theta\bigg]
  \bigg[2\left(\int_{0}^{\pi/2} - \int_{\pi/2}^{\pi} \right)d\chi\bigg].
\end{eqnarray}
We denote the two asymmetries $A_{\rm con}$ and $A_{\rm Pcon}$ associated with
the square of $\cos\theta$ as the unpolarized and polarized convexity
asymmetries because the respective angular decay distributions are described
by a tilted upward or downward paraboloid depending on the sign of the
corresponding coefficients ${\cal O}_2$ and ${\cal O}_5$. The convexity
coefficients could also be isolated by taking the second derivative of the
angular decay distribution w.r.t.\ $\cos\theta$.

Corresponding to the total and partial differential rates~(\ref{q2difftot})
and~(\ref{q2partial}) we define integrated total and partial rates according
to 
\begin{equation}
\Gamma_{\rm tot} = \int_{m_e^2}^{(M_n^2-M_p^2) }dq^2
  \frac{d\Gamma_{\rm tot}}{dq^2},\qquad \qquad \qquad
\Gamma_i = \int_{m_e^2}^{(M_n^2-M_p^2)}dq^2 \frac{d\Gamma_i}{dq^2}.
\end{equation}
This leads us to the definition of the average values of the observables
$\langle {\cal O}_i \rangle$ which are the focus of the analysis in our paper.
One has
\begin{equation}
\langle {\cal O}_i \rangle = \frac{\Gamma_i}{\Gamma_{\rm tot}}
  = \frac{\int_{m_e^2}^{(M_n^2-M_p^2) }dq^2\,F(q^2){\cal H}_i(q^2) }
  {\int_{m_e^2}^{(M_n^2-M_p^2) }dq^2F(q^2){\cal H}_{\rm tot}(q^2)},
\end{equation}
where $F(q^2)$ is a $q^2$-dependent phase-space factor given by
$F(q^2)=(q^2-m_e^2)^2p/q^2$.

\section{Unpolarized neutron decays\label{sec4}}
We begin our discussion with the total differential rate given by
\begin{eqnarray}
\label{diffrateq^2}
\frac{d\Gamma_{\rm tot}}{dq^2}
  &=& \frac{\Gamma_0(q^2-m_e^2)^2 p}{M_n^7q^2} \;
  \Bigg[ {\cal H}_U + {\cal H}_L +\; \delta_e \;
  \Big\{ 3 {\cal H}_{S} + {\cal H}_U + {\cal H}_L \Big\} \;\Bigg]\;.  
\end{eqnarray}
As mentioned before the squared momentum transfer can be determined from the
energy or momentum of the decay proton in the neutron rest frame. Note that
ultracold neutrons (UCN) are practically at rest when they decay. They have
typical velocities of 5~ms$^{-1}$ which corresponds to a kinetic energy of the
neutron of $E_e^{\rm kin}=E_e-m_n=1.31\times 10^{-4}\MeV$.

In Eq.~(\ref{diffrateq^2}) we have separated the helicity nonflip and helicity
flip contributions where the last three terms in Eq.~(\ref{diffrateq^2})
muliplied by the helicity flip factor $\delta_e= m_e^2/(2q^2)$ represent the
helicity flip contribution. In Fig.~\ref{dgamma} we present a plot of the
$q^2$ dependence of the partial differential helicity nonflip rate
$d\Gamma_{\rm nf}/dq^2$ and helicity flip rate $d\Gamma_{\rm hf}/dq^2$ as well
as the sum of the two which is nothing but the total differential rate. The
helicity flip contributions are negligibly small for $\ell= e,\mu$ in
semileptonic bottom hadron decays and quite small (${\cal O}(1\,\%)$) for
$\ell=\mu$ in semileptonic charm hadron decays. Contrary to this, the helicity
flip contribution can obviously not be neglected in neutron $\beta$ decay. The
helicity flip factor $\delta_e=m_e^2/(2q^2)$ can become quite large close to
threshold $q^2=m_e^2$. Numerically one has\qquad $ 0.5\ge \delta_e \ge0.07805$
and $0.7071\ge \sqrt{\delta_e} \ge0.2794 $. We have also listed the
corresponding range of $\sqrt{\delta_e}$ which determines the flip suppression
for the transverse polarization of the electron to be discussed later on. The
total rate as well as the partial rates vanish at threshold $q^2=m_e^2$
(maximal recoil) and at zero recoil $q^2=(M_n-M_p)^2$ due to the overall
kinematical factors $(q^2-m^2_e)^2$ and $p \sim ((M_n-M_p)^2-q^2)^{1/2}$ in
the rate expression~(\ref{angdist1}). At zero recoil one has
$d\Gamma_{\rm hf}/d\Gamma_{\rm nf}=(1+(F_1^V/F_1^A)^2)\delta_e(\rm z.r.)$
where the zero-recoil value of $\delta_e$ is $\delta_e(\rm z.r.)=0.078$ (see
above). The total differential rate can be seen to be almost symmetrically
distributed w.r.t.\ to its peak position at around $q^2=1\MeV^2$.

\begin{figure}\begin{center}
\epsfig{figure=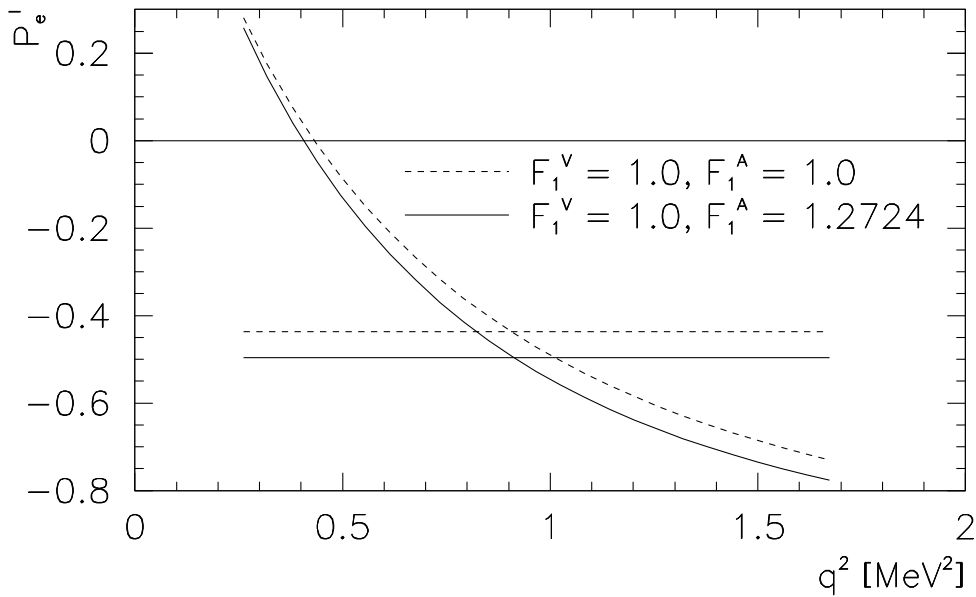, scale=0.8}
\caption{\label{poll}$q^2$ dependence of the longitudinal polarisation
  $P_e^\ell(q^2)$ for two different sets of form factors (solid and dashed
  lines). The straight lines represent the longitudinal polarisation
  integrated over $q^2\in[m_\ell^2,(M_n-M_p)^2]$.}
\end{center}\end{figure}

The separation into helicity nonflip and flip contributions allows one to
immediately conclude for the average longitudinal polarization of the electron
in the $q$ frame along the electron's momentum direction when one has
integrated over the the three correlation angles $(\theta,\,\theta_P,\,\chi)$.
One has
\begin{equation}
P_e^\ell=\frac{d\Gamma_{\rm hf}-d\Gamma_{\rm nf}}
  {d\Gamma_{\rm hf}+d\Gamma_{\rm nf}},
\end{equation}
since the helicity nonflip and flip rates correspond to the electron's
helicity values of $\lambda_e=-1/2$ and $\lambda_e=+1/2$, respectively.
Close to threshold $q^2=m_e^2$ where the differential rates are small, the
flip contribution is larger than the nonflip contribution in a narrow range of
$q^2$ implying a small positive value of the longitudinal polarization of the
electron, as can be seen in Fig.~\ref{poll} where we plot the $q^2$ dependence
of the longitudinal polarization of the electron. Starting at around
$q^2=0.5\,{\rm MeV^2}$ the differential rate is dominated by the nonflip
contribution such that the longitudinal polarization of the electron is
negative in the remaining $q^2$ range as again evidenced in Fig.~\ref{poll}.
The longitudinal polarization at zero recoil is determined by the flip/nonflip
ratio at zero recoil calculated above which results in $P_e^\ell=-0.776$ at
zero recoil in agreement with Fig.~\ref{poll}. The average value of the
longitudinal polarization $\langle P_e^\ell \rangle \approx 0.50$ is close to
the value of the polarization $P_e^\ell(q^2)$ at the peak position of the
differential rate at around $q^2\approx 1\GeV^2$. 
In Sec.~\ref{sec6} we shall also present results on the transverse
polarization of the decay electron.

\begin{figure}\begin{center}
\epsfig{figure=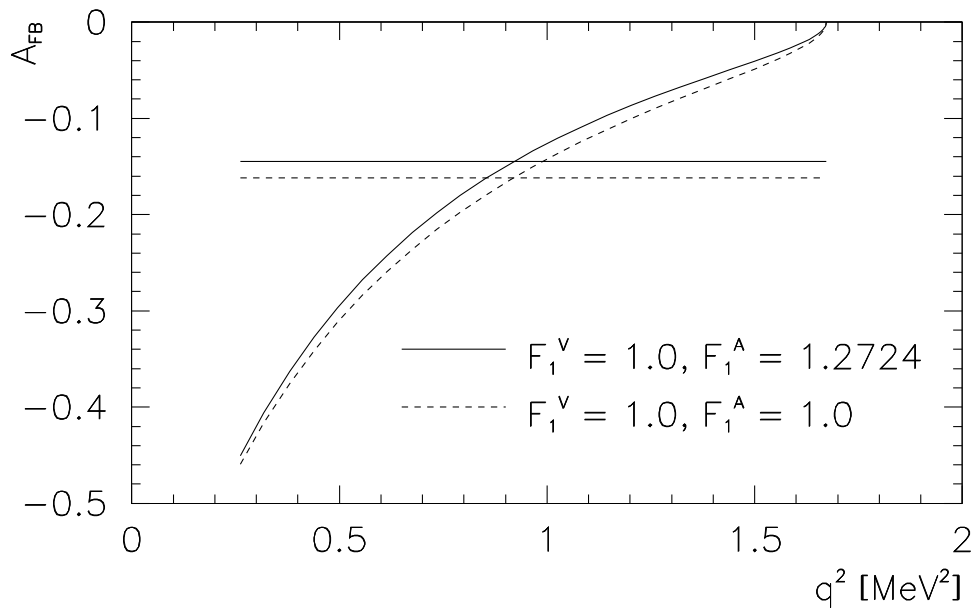, scale=0.8}
\caption{\label{afb1}Forward--backward asymmetry $A_{\rm FB}(q^2)$ as a
  function of $q^2$ for two different sets of form factors (solid and dashed
  lines). The straight lines represent the average value forward--backward
  asymmetry integrated over $q^2\in[m_\ell^2,(M_n-M_p)^2]$.}
\end{center}\end{figure}

In Fig.~\ref{afb1} we present a plot of the $q^2$ dependence of the
forward--backward asymmetry $A_{\rm FB}(q^2)$ where, according to
Eq.~(\ref{FB1}), $A_{\rm FB}(q^2)$ is given by
\begin{equation}
A_{\rm FB}(q^2) = \frac{1}{2} \frac{{\cal H}_1(q^2)}{{\cal H}_{\rm tot}(q^2)}.
\end{equation}
The forward--backward asymmetry starts with a rather large negative value
$A_{\rm FB}(m_e^2)=-0.46$ at threshold and goes to zero at zero recoil. The
vanishing of the forward--backward asymmetry at zero recoil can be understood
from the fact that ${\cal H}_1$ is proportional to
$( {\cal H}_F + 4\delta_e{\cal H}_{SL_+})$. Both of these components vanish in
the zero-recoil limit (see Eq.~(\ref{eq:zerec})). 

If there is enough data and if the energy of the recoiling proton can be
measured, it would certainly be interesting to take a detailed look at the
$q^2$ dependence of the rate and the various polarization observables. In a
more inclusive analysis one can also consider $q^2$-integrated quantities such
as the total rate and the $q^2$ averages of the various polarization
observables. It turns out that the $q^2$ integration of the rate and the
polarization observables can be done analytically even including possible
$q^2$ dependencies of the form factors. However, the resulting analytical
expressions become quite long and unwieldy. A much more transparent and
discerning representation of the integrated quantities can be obtained by
performing a recoil expansion of the analytical results in terms of powers
of the small parameter
\begin{equation}
\delta=(M_n-M_p)/(M_n+M_p)= 0.689\cdot 10^{-3}. 
\end{equation}
In the recoil expansion it is convenient to split off an overall factor of
$1/(1+\delta)^8$. We thus write
\begin{equation}
\Gamma_i=\frac{1}{(1+\delta)^8}\left( \Gamma_i^{(5)}\delta^5
+\Gamma_i^{(6)}\delta^6 +\ldots \right)
\end{equation}
and formally call $\Gamma_i^{(5)}$, $\Gamma_i^{(6)}$ or common constant
fractions of them the LO, NLO contributions in the recoil expansion.

In order to exhibit form factor effects and the linear contributions of the
form factor $F^A_3$, the recoil expansion has to be done up to NNLO order,
namely up to the order ${\cal O}(\delta^7)$, where these contributions first
appear. For the recoil expansion of the total rate one obtains
\begin{eqnarray}
 \lefteqn{\Gamma\ =\ \frac{512\Gamma_0}{5(1+\delta)^8}
  \times\Big[\Big\{\left(3(F_1^A)^2+(F_1^V)^2\right)\,r(x)\Big\}\,\delta^5
  \strut}\nonumber\\&&\strut\kern-12pt
  +\Big\{5F_1^VF_3^Vx^2\left((2+13x^2)\sqrt{1-x^2}+3x^2(4+x^2)
  L_2\right)\strut\nonumber\\&&\strut
  -F_1^AF_2^A\left((8-26x^2+33x^4)\sqrt{1-x^2}+15x^6
  L(x)\right)\Big\}\delta^6
  \strut\nonumber\\&&\strut\kern-12pt
  +\Big\{\frac{48}{7}(F_2^A)^2(1-x^2)^{7/2}\strut\nonumber\\&&\strut
  +2(F_3^V)^2x^2\left((2-9x^2-8x^4)\sqrt{1-x^2}-15x^4 L(x)\right)
  \strut\nonumber\\&&\strut
  +F_1^AF_2^A\left((8-26x^2+33x^4)\sqrt{1-x^2}+15x^6 L(x)\right)
  \strut\nonumber\\&&\strut
  +\frac{1}{7}(3F_1^V+2F_2^V)F_2^V\left((8-38x^2+87x^4+48x^6)\sqrt{1-x^2}
  +105x^6L(x)\right)\strut\nonumber\\&&\strut
  -5F_1^VF_3^Vx^2\left((2+13x^2)\sqrt{1-x^2}+3x^2(4+x^2)L(x)\right)
  \strut\nonumber\\&&\strut
  -F_1^AF_3^Ax^2\left((6+83x^2+16x^4)\sqrt{1-x^2}+15x^2(4+3x^2)L(x)\right)
  \strut\nonumber\\&&\strut
  -\frac{1}{14}(F_1^A)^2\left((30-69x^2+188x^4-44x^6)\sqrt{1-x^2}
  +105x^4L(x)\right)\strut\nonumber\\&&\strut
  +\frac{3}{14}(F_1^V)^2
  \left((2-27x^2-92x^4+12x^6)\sqrt{1-x^2}-105x^4L(x)\right)
  \strut\nonumber\\&&\strut
  +\frac{2}{21}(F_1^A)^2M_n^2\left((20-32x^2+319x^4+8x^6)\sqrt{1-x^2}
  +105x^4(2+x^2)L(x)\right)\langle(r_1^A)^2\rangle
  \strut\nonumber\\&&\strut
  +\frac{2}{21}(F_1^V)^2M_n^2\left((4+16x^2+271x^4+24x^6)\sqrt{1-x^2}
  +105x^4(2+x^2)L(x)\right)\langle(r_1^V)^2\rangle\Big\}\,\delta^7
  +\ldots\Big].\qquad
\label{nnlorate}
\end{eqnarray}
Since we assume the form factors to be relatively real, we use a simplified
notation and write $(F_1^A)^2=|F_1^A|^2$,
$F_1^A\,F_2^A=\real(F_1^AF_2^{A\dagger})$ etc.\ in Eq.~(\ref{nnlorate}) and
elsewhere. The LO ${\cal O}(\delta^5)$ contribution has the familiar form
proportional to $\left(3(F_1^A)^2+(F_1^V)^2\right)r(x)$ where ($r(0)=1$)
\begin{equation}
\label{rofx}
r(x)=\frac12\left((2-9x^2-8x^4)\sqrt{1-x^2}
  -15x^4\ln\left(\frac{1-\sqrt{1-x^2}}{x}\right)\right)=0.4726,
\end{equation}
and where
\begin{equation}
x=m_e/(M_n-M_p)=0.395\,.
\end{equation}
It is quite remarkable that the electron mass dependence factors out in the LO
term. If the second class form factor contributions are set to zero, the NLO
contribution in the formal recoil expansion can be seen to vanish. This has
been noted before in Refs.~\cite{Kadeer:2005aq,Chang:2014iba}. The NLO
contribution is again proportional to $\left(3(F_1^A)^2+(F_1^V)^2\right)r(x)$
in a formal sense if one expands $1/(1+\delta)^8=1-8\delta+\ldots$\,.

As Eq.~(\ref{nnlorate}) shows, the factorization of the electron mass
dependence is no longer true for the higher order terms in the recoil
expansion. The higher order terms in Eq.~(\ref{nnlorate}) contain the
logarithmic factor
\begin{equation}
L(x):=\ln\pfrac{1-\sqrt{1-x^2}}{x}=-1.580
\end{equation}
which is always multiplied by powers of $x$ such that the $L(x)$ contribution
vanishes in the zero electron mass limit.

In the literature the phase space factor needed for the total rate is usually
obtained by first integrating over $E_\nu$ to obtain the $E_e$ spectrum
followed by the integration over the $E_e$ spectrum. One obtains the
well-known simple expressions only if one introduces zero recoil
approximations in the integrand from the very beginning. Without zero recoil
approximations the integration over $(E_\nu,E_e)$ phase space becomes quite
complicated even in the unpolarized case (see Ref.~\cite{Bender:1968zz}
and Ref.~\cite{Wilkinson:1982hu} Sec.~15.1) and is best done numerically.
Compare this to our integration over the $(\cos\theta,q^2)$ phase space which
is quite straightforward because the phase-space integrations factorize. In
addition, the first $\cos\theta$ integration is trivial. The $(\cos\theta,q^2)$
integration route allows one to obtain compact expressions for the
coefficients of the recoil expansion, as Eq.~(\ref{nnlorate}) shows. We
emphasize that we do the zero recoil expansion after having done the full
integration whereas in the $(E_\nu,E_e)$ phase space calculations the recoil
expansion is frequently done prior to the last $E_e$
integration~\cite{Wilkinson:1982hu} which may lead to inaccurate results. 

The contribution of the large induced pseudoscalar form factor $F_3^A$ sets in
only at NNLO in the recoil expansion where it enters linearly. It is
multiplied by the $x$-dependent factor
\begin{equation}
x^2\left((6+83x^2+16x^4)\sqrt{1-x^2}+15x^2(4+3x^2)L(x)\right)=0.193.
\end{equation}
The accompanying relative recoil factor $\delta^2=0.475\cdot10^{-6}$ reduces
the linear ${\cal O}(\delta^7)$ rate contribution of $F_3^A$ to an
insignificant level. As it turns out, the same observation is true for the
contribution of $F_3^A$ to all other partial rates.

In order to check on the sensitivity of the recoil expansion to the $q^2$
dependence of the form factors we have made a linear Ansatz for the form
factors in terms of the isovector radii, i.e.\ we write
\begin{equation}
F_i^X(q^2)=F_i^X(0)\left(1+\langle(r_i^X)^2\rangle \frac{q^2}{6}\right)
\end{equation}
for $i=1,2,3$; $X=V,A$. Eq.~(\ref{nnlorate}) shows that the form factor
dependence of the form factors $F_1^{V/A}(q^2)$ sets in only at NNLO while the
form factor dependence of $F_{2,3}^{V/A}(q^2)$ contributes only to higher
orders in the recoil expansion. For the radii of the $F_1^V(q^2)$ and
$F_1^A(q^2)$ form factors we take
$\langle(r_1^V)^2\rangle=0.66\,{\rm fm^2}=1.695\times 10^{-5} {\rm MeV^{-2}}$
and
$\langle(r_1^A)^2\rangle=0.45\,{\rm fm^2}=1.156\times 10^{-5} {\rm MeV^{-2}}$
\cite{Bourquin:1981ba,Faessler:2008ix}.

In order to be able to assess the importance of the $q^2$ dependence of the
form factors in the ${\cal O}(\delta^7)$ terms we take a closer numerical look
at the first LO term and the last four ${\cal O}(\delta^7)$ terms in
Eq.~(\ref{nnlorate}). One has
\begin{eqnarray}
\label{numradii}
\Gamma_{\rm tot}&=&\frac{512\Gamma_0\,\delta^5}{5(1+\delta)^8}
\Big[\Big\{\Big(3(F_1^A)^2+(F_1^V)^2\Big)0.47\Big\}\ldots \nn &&
  -\Big\{\frac{1}{14}(F_1^A)^2 \Big(9.245-166.423\Big)
  +\frac{3}{14}(F_1^V)^2 \Big(12.782-22.623\Big)\Big\}\,\delta^2\,+\dots\Big],
\end{eqnarray}
where the numbers $166.423$ and $22.623$ refer to the
$\langle(r_1^A)^2\rangle$ and $\langle(r_1^V)^2\rangle$ contributions.
Eq.~(\ref{numradii}) shows that the $q^2$-dependent NNLO form factor
contributions can become quite large compared to their $q^2$-independent
NNLO counterparts. However, when multiplied by $\delta^2=0.475\cdot10^{-6}$
the overall contribution of the $q^2$-dependent NNLO form factor contributions
is insignificant. We have checked that this is true for all partial rates
treated in this paper.

Returning to Eq.~(\ref{nnlorate}) one notes the remarkable result that the
${\cal O}(\delta^6)$ term in the rate expansion vanishes altogether when the
second class current contributions are set to zero, i.e.\ for $F_3^V=F_2^A=0$.
The second class currents can be expected to be at most of
${\cal O}(\delta^1)$, and thus the initial NLO ${\cal O}(\delta^6)$
contribution of the second class form factors would be shifted up to the order
${\cal O}(\delta^7)$. In fact, in a $SU(6)_W$ quark model calculation one finds
$F_3^V\approx (M_n-M_p)$ and $F_2^A=0$~\cite{Hussain:1990ai}. The absence of
NLO contributions in the recoil expansion of the rate implies that the NLO
corrections to the $q^2$ average of an observable are entirely determined by
the NLO correction to the partial rate associated with the observable.

This can be seen as follows. Consider the $q^2$ average
$\langle{\cal O}_i\rangle$ of a given observable ${\cal O}_i$. In the recoil
expansion one has
\begin{equation}
\label{recexp}
\langle {\cal O}_i\rangle
  =\frac{(\Gamma_i^{(5)}\,\delta^5 + \Gamma_i^{(6)}\,\delta^6 + \ldots)}
  {(\Gamma_{\rm tot}^{(5)}\, \delta^5 + \Gamma_{\rm tot}^{(7)}\,\delta^7 
  + \ldots)}
  =\frac{\Gamma_i^{(5)}}{\Gamma_{\rm tot}^{(5)}}\,\,\bigg(1
  +\Gamma_i^{(6)}/\Gamma_i^{(5)}\,\delta+\Big(\Gamma_i^{(7)}/\Gamma_i^{(5)}
  -\Gamma_{\rm tot}^{(7)}/\Gamma_{\rm tot}^{(5)}\Big)\,\delta^2+\ldots\bigg),
\end{equation}
which shows that the NLO correction to $\langle{\cal O}_i\rangle$ is solely
determined by the ratio $\Gamma_i^{(6)}/\Gamma_i^{(5)}$ when the contributions
of the second class currents are set to zero. In Sec.~\ref{sec7} we provide
numerical results on the LO ratios $\Gamma_i^{(5)}/\Gamma_{\rm tot}^{(5)}$ as
well as the NLO corrections $\Gamma_i^{(6)}/\Gamma_i^{(5)}$ for the various
observables.

We now list the recoil expansion for the two unpolarized partial rates
$\Gamma_{\rm FB}$ and $\Gamma_{\rm conv}$. One has
\begin{itemize}
\item p.v.\ partial forward--backward rate $\Gamma_{\rm FB}$ \quad
  (${\cal H}_{\rm FB}=\frac 12\,{\cal H}_1 = -\frac 34\,({\cal H}_F
  +4\delta_e {\cal H}_{SL_+}$)
\begin{eqnarray}
\label{FB}
\Gamma_{\rm FB} &=&\frac 12 \,\,\Gamma_1
 \ =\ \frac{64\Gamma_0}{(1+\delta)^8}
  \Big[3x^2((F_1^V)^2+(F_1^A)^2)\left((1-x^2)(5+x^2)+4(1+2x^2)\ln x\right)
  \delta^5\strut\nonumber\\&&\strut
  +2\Big((1-x^2)\left((F_1^V+2F_2^V)F_1^A(1-5x^2-2x^4)
  -3(F_1^VF_3^V-F_1^AF_2^A)x^2(1+5x^2)\right)\strut\nonumber\\&&\strut\qquad
  -12x^4\left((F_1^V+2F_2^V)F_1^A+(F_1^VF_3^V-F_1^AF_2^A)(2+x^2)\right)\ln x
  \Big)\delta^6+O(\delta^7)\Big].
\end{eqnarray}
The LO term in the recoil expansion can be seen to be entirely given by
the longitudinal--scalar interference term $\delta_e{\cal H}_{SL_+}$ with the
characteristic overall factor $x^2\sim m_e^2$. The parity-violating structure
function ${\cal H}_F$ comes in only at NLO and is proportional to
$(F_1^V+2F_2^V)F_1^A$ when $F_3^V=F_2^A=0$. 

\item p.c.\ partial convexity rate $\Gamma_{\rm conv}$ \quad
(${\cal H}_{\rm conv}=\frac 38\,{\cal H}_2
=\frac{3}{16}(1-2\delta_e) ({\cal H}_U-2{\cal H}_L$)) \\ 
For the integrated partial rate $\Gamma_{\rm conv}$ associated with the
$\cos^2\theta$ contribution one obtains the recoil expansion
\begin{eqnarray}
\label{conv}
\Gamma_{\rm conv}&=& \frac 38 \Gamma_2
  \ =\  -\frac{24\Gamma_0}{5(1+\delta)^8}
  \Bigg[7\left((F_1^V)^2+(F_1^A)^2\right)\Bigg((8+194x^2+113x^4)\sqrt{1-x^2}
  \strut \nn &&\strut\kern-24pt
  +15x^2(8+12x^2+x^4)\,L(x)\Bigg)\delta^5
  -2\left(3(F_1^V)^2+3(F_1^A)^2+8(F_2^V)^2+8(F_2^A)^2\right)
  \times\kern-4pt\strut\nonumber\\&&\strut\kern-12pt\times
  \Bigg((4-40x^2-247x^4-32x^6)\sqrt{1-x^2}
  -105x^4(2+x^2)\,L(x)\Bigg)\delta^7+O(\delta^8)\Bigg].
\end{eqnarray}
One notes that their are no NLO contributions to the p.c.\ partial rate
$\Gamma_{\rm conv}$ when $F_3^V=F_2^A=0$.
\end{itemize}
\section{Polarized neutron decays\label{sec5}}
With the availability of polarized neutron sources the number of possible
correlation measurements in neutron $\beta$ decays is increased from two to
seven as Eq.~(\ref{angdist4}) shows. The neutron spin correlation measurements
are proportional to the magnitude of the polarization of the neutron $P_n$,
the value of which needs to be known to a high accuracy. Fortunately, one
can presently avail of neutron beams with a very high degree of polarization
close to 100$\%$~\cite{Surkau:1997,Kreuz:2005,Brown:2017mhw}.

We now list the partial rates needed for the numerators of the five polarized
obserables $\langle{\cal O}_3\rangle$ to $\langle {\cal O}_7\rangle$ where we
include the respective projection factors from Eq.~(\ref{FB2}). The recoil
expansion is carried out up to NLO. One obtains
\begin{itemize}
\item p.v.\ polarized forward--backward asymmetry $\Gamma_{\rm PFB}$ \quad
  (${\cal H}_{\rm PFB}(q^2)=\frac 12 {\cal H}_3(q^2)=-\frac 12(1+\delta_e)
  ({\cal H}_F-{\cal H}_{L_-})+\frac 32 \delta_e {\cal H}_{S_-}$)
\begin{eqnarray}
\label{PFB}
\Gamma_{\rm PFB} &=& \frac{128\Gamma_0}{3(1+\delta)^8}
  \Big[3F_1^VF_1^A\left((1-x^2)(1-5x^2-2x^4)-12x^4\ln x\right)\delta^5
  \strut\nonumber\\&&\strut
  -\Big((1-x^2)\left((2-7x^2+11x^4)\left((F_1^V+2F_2^V)F_1^A+F_1^VF_2^A\right)
  -9x^2(1+5x^2)F_3^VF_1^A\right)\strut\nonumber\\&&\strut\qquad
  +12x^4\left(x^2\left((F_1^V+2F_2^V)F_1^A+F_1^VF_2^A\right)
  -3(2+x^2)F_1^AF_3^V\right)\ln x\Big)\delta^6+O(\delta^7)\Big].
\end{eqnarray}
For $F_3^V=F_2^A=0$ the axial form factor $F_1^A$ factors out and one arrives
at the simple form
\begin{eqnarray}
\Gamma_{\rm PFB} &=& \frac{128\Gamma_0}{15(1+\delta)^8}\,F_1^A
  \Big[3F_1^V\left((1-x^2)(1-5x^2-2x^4)-12x^4\ln x\right)\delta^5
    \strut\nonumber\\&&\strut
    +(F_1^V+2F_2^V)\Big(2-9x^2 +18x^4+x^6\Big)\delta^6 +O(\delta^7)\Big]
  \end{eqnarray}
The LO contribution agrees with the corresponding result of
Ref.~\cite{Gudkov:2008pf}.

\item p.c.\ double forward--backward asymmetry \quad
  (${\cal H}_{\rm DFB}(q^2)=\frac 14 {\cal H}_4(q^2)=\frac 38 ({\cal H}_U -
  4\delta_e {\cal H}_{SL_-})$)
\begin{eqnarray}
\label{DFB}
\Gamma_{\rm DFB} &=& \frac{32\Gamma_0}{5(1+\delta)^8}\Bigg[F_1^A
  \Bigg(\left(5x^2(46+29x^2)F_1^V-4(2-9x^2-8x^4)F_1^A\right)\sqrt{1-x^2}
  \strut\nonumber\\&&\strut\qquad
  +15x^2\left((8+16x^2+x^4)F_1^V+4x^2F_1^A\right)\,L(x)
  \Bigg)\delta^5\strut\nonumber\\&&\strut
  +4\Bigg(\left(5x^2(2+13x^2)(F_1^VF_2^A-F_3^VF_1^A)
  +4(2-9x^2-8x^4)F_1^AF_2^A\right)\sqrt{1-x^2}\strut\nonumber\\&&\strut\qquad
  +15x^4\left((4+x^2)(F_1^VF_2^A-F_3^VF_1^A)
  -4F_1^AF_2^A\right)L(x)\Bigg)\delta^6
  +O(\delta^7)\Bigg]\,.
\end{eqnarray}
The NLO contribution can be seen to vanish for $F_3^V=F_2^A=0$.
\item p.v.\ polarized convexity parameter \quad
  (${\cal H}_{\rm Pconv}(q^2)=\frac{3}{16} {\cal H}_5(q^2)
  =-\frac{3}{32}(1-2\delta_e)({\cal H}_F+2{\cal H}_{L_-})$)
\begin{eqnarray}
\label{Pconv}
\Gamma_{\rm Pconv} &=& \frac{-16\Gamma_0}{(1+\delta)^8}\Big[3F_1^VF_1^A
  \left((1-x^2)(1+10x^2+x^4)+12x^2(1+x^2)\ln x\right)\delta^5
  \strut\nonumber\\&&\strut
  +\left((F_1^V+2F_2^V)F_1^A-2F_1^VF_2^A\right)
  \Big((1-x^2)(1-8x^2-17x^4)\strut\nonumber\\&&\strut\qquad
  -12x^4(3+x^2)\ln x\Big)\delta^6+O(\delta^7)\Big]\,,
\end{eqnarray}
\item p.c.\ azimuthal asymmetry 1 \quad
  (${\cal H}_{\chi_1}(q^2)=\frac{\pi}{8} {\cal H}_6(q^2)
  =-\frac{3\pi}{8\sqrt{2}}({\cal H}_{LT_+}-2\delta_e{\cal H}_{ST_-})$)
\begin{eqnarray}
\label{chi1}
\Gamma_{\chi_1} &=& \frac{-4\pi\Gamma_0}{(1+\delta)^8}\bigg[2F_1^A
  \Big(x\sqrt{1-x^2}\left(4x^2(13+2x^2)F_1^V-3(1+14x^2)F_1^A\right)
  \strut\nonumber\\&&\strut\qquad
  -3\left(4x^2(1+4x^2)F_1^V+(1-8x^2-8x^4)F_1^A\right)\arccos(x)\Big)\delta^5
  \strut\nonumber\\&&\strut
  -\bigg(x\sqrt{1-x^2}\Big((3+94x^2+8x^4)F_1^V(F_1^V+2F_2^V)
  -10x^2(11+10x^2)(F_1^V+2F_2^V)F_1^A\strut\nonumber\\&&\strut\qquad\quad
  -2(9+74x^2-8x^4)F_1^AF_2^A+16x^2(13+2x^2)F_1^VF_2^A+12x^2(1+14x^2)F_3^VF_1^A
  \Big)\strut\nonumber\\&&\strut\qquad
  +3\Big((1-12x^2-24x^4)F_1^V(F_1^V+2F_2^V)
  +2x^2(3+24x^2+8x^4)(F_1^V+2F_2^V)F_1^A\strut\nonumber\\&&\strut\qquad\qquad
  -2(3-20x^2-8x^4)F_1^AF_2^A-16x^2(1+4x^2)F_1^VF_2^A
  \strut\nonumber\\&&\strut\qquad\qquad
  +4x^2(1-8x^2-8x^4)F_3^VF_1^A\Big)\arccos(x)\bigg)\delta^6+O(\delta^7)\bigg],
\end{eqnarray}
\item p.v.\ azimuthal asymmetry 2 \quad
  (${\cal H}_{\chi_2}(q^2)=\frac 13 {\cal H}_7(q^2)
  =\frac{1}{2\sqrt{2}}(1-2\delta_e){\cal H}_{LT_-}$)
\begin{eqnarray}
\label{chi2}
\Gamma_{\chi_2} &=& \frac{512\Gamma_0}{15(1+\delta)^8}
  (1-x)^5(1+5x)\Big[F_1^VF_1^A\delta^5\strut\nonumber\\&&\strut
  +\left((F_1^V+2F_2^V)F_1^A-2F_1^VF_2^A\right)\delta^6+O(\delta^7)\Big].\nn
\end{eqnarray}
\end{itemize}

\section{Polarization of the decay electron\label{sec6}}
In Sec.~\ref{sec4} we have already discussed some aspects of the longitudinal
polarization of the decay electron. In this section we provide explicit LO and
NLO expressions needed for the calculation of the average of the longitudinal
polarization $\langle P_e^\ell\rangle$. We also extend the discussion to the
transverse component of the decay electron. In all generality the two
polarization components depend on the correlation angles
$(\theta,\theta_P,\chi)$. In this work we consider only averages of the two
polarization components where the averaging is done w.r.t.\ the three
correlation angles $(\theta,\theta_P,\chi)$. This implies that we do not
consider the correlation of the electron polarization with the neutron
polarization as has been done e.g.\ in Ref.~\cite{Nico:2009zua}.

Using a slightly modified version of the master formula~(\ref{master1}) one
can calculate the differential $q^2$ distributions of the numerators of the
relevant polarization expressions. One has
\begin{eqnarray}
\label{epoll1}
\frac{d\Gamma(P_e^\ell)}{dq^2}
  &=& \frac{d\Gamma_{\rm hf}}{dq^2}-\frac{d\Gamma_{\rm nf}}{dq^2}
  \ =\ \frac{\Gamma_0(q^2-m_e^2)^2p}{M_n^7q^2}\bigg(\delta_e
  \Big(3{\cal H}_S+{\cal H}_U+{\cal H}_L\Big)
  -\Big({\cal H}_U+{\cal H}_L\Big)\bigg),\qquad \mbox{(p.c.)} \\
\label{epolt1}
  \frac{d\Gamma(P_e^t)}{dq^2}
  &=& 2\frac{d\Gamma(\lambda_e=\tfrac 12,\lambda'_e=-\tfrac 12)}{dq^2}
  \ =\ -\frac{3\pi}{4}\frac{\Gamma_0(q^2-m_e^2)^2p}{M_n^7q^2}
  \sqrt{\frac{\delta_e}{2}} \Big({\cal H}_F -2{\cal H}_{SL_+}\Big).
  \qquad \qquad \mbox{(p.v.)}
\end{eqnarray}
The transverse polarization is proportional to the interference of the
nonflip and flip helicity amplitudes and is thus proportional to the square
root $\sqrt{\delta_e}$ of the helicity flip penalty factor. One needs to know
the sign of the interference contribution which is given by
\begin{equation}
h_{\lambda_e=1/2\,\lambda_{\nu}=1/2} / \,h_{\lambda_e
  = -1/2\,\,\lambda_{\nu}=1/2} = \sqrt{m_e^2/2q^2} = \sqrt{\delta_e}.
\end{equation}

\begin{figure}\begin{center}
\epsfig{figure=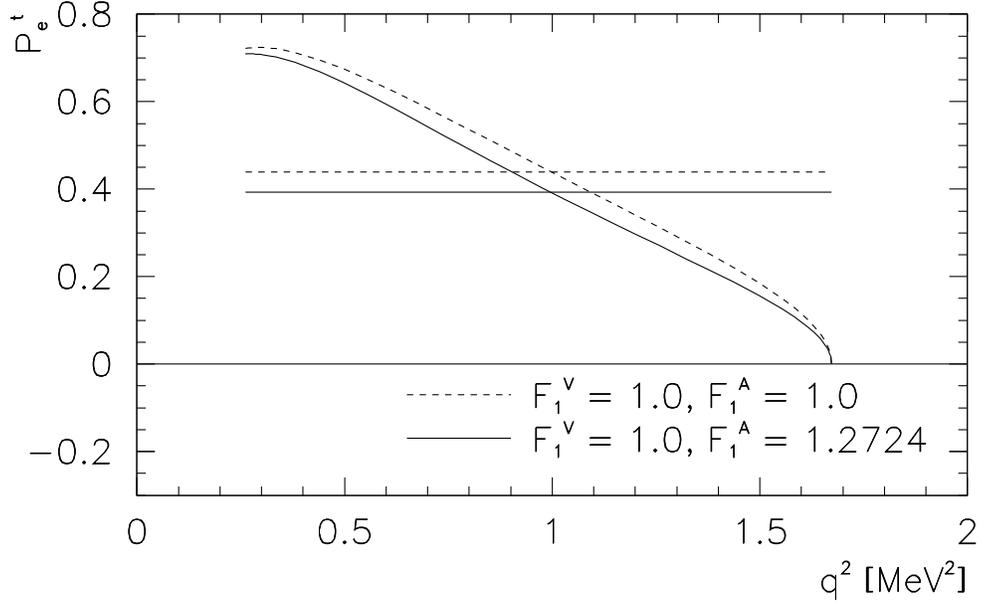, scale=0.8}
\caption{\label{polt}Transverse polarisation of the electron as a function of
  $q^2$ for two different sets of form factors (solid and dashed lines). The
  straight lines represent the longitudinal polarisation integrated over
  $q^2\in[m_\ell^2,(M_n-M_p)^2]$.}
\end{center}\end{figure}

The corresponding expressions for the two components of the polarization are
given by
\begin{equation}
\label{epol}
P_e^\ell = \frac{d\Gamma(P_e^\ell)}{dq^2}\Big/ \frac{d\Gamma_{\rm tot}}{dq^2}
  \qquad
  P_e^t = \frac{d\Gamma(P_e^t)}{dq^2}\Big/ \frac{d\Gamma_{\rm tot}}{dq^2}
\end{equation}
As expected, the electron can be seen to be $100\,\%$ longitudinally
polarized $P_e^\ell=-1$ in the limit of a vanishing electron mass, i.e.\ when
setting $\delta_e=0$ in Eq.~(\ref{epoll1}). In the same limit the transverse
component vanishes as can again be seen by setting $\sqrt{\delta_e}=0$ in
Eq.~(\ref{epolt1}). In Fig.~\ref{polt} we show a plot of the transverse
polarization of the electron. The transverse polarization starts with a rather
large positive value at threshold and then drops to zero at zero recoil. The
vanishing at zero-recoil results from the fact that both ${\cal H}_F$ and
${\cal H}_{SL_+}$ vanish at zero recoil (see Eq.~(\ref{eq:zerec})). As in the
case of the longitudinal polarization of the electron, the average value
$\langle P_e^t\rangle \approx 0.45$ of the transverse polarization is close to
the value of $P_e^t(q^2)$ at the peak position of the differential total rate. 

Next we integrate the numerators of the two polarization components in
Eq.~(\ref{epol}) over $q^2$ and expand the resulting expressions up to NLO in
the recoil parameter $\delta$. One has
\begin{eqnarray}
\label{epoll2}
\Gamma(P_e^\ell)&=&-\frac{256\Gamma_0}{15(1+\delta)^8}
  \Bigg\{\Bigg[\left((F_1^V)^2r(x)+(F_1^A)^2 r(x)\right)
  \Bigg]\delta^5\strut\nonumber\\&&\strut
  -6\Bigg[\left(5x^2(2+13x^2)F_1^VF_3^V+(8-46x^2-97x^4)F_1^AF_2^A\right)
  \sqrt{1-x^2}\strut\nonumber\\&&\strut\qquad
  +15x^4\left((4+x^2)F_1^VF_3^V-(8+x^2)F_1^AF_2^A\right)L(x)
  \Bigg]\delta^6+O(\delta^7)\Bigg\},\\
\label{epolt}
\Gamma(P_e^t)&=&\frac{256\pi\Gamma_0}{(1+\delta)^8}x(1-x)^4
  \Bigg\{\left[(F_1^V)^2+(F_1^A)^2\right]\delta^5\strut\nonumber\\&&\strut
  +\frac25(1+4x)\left[F_1^VF_3^V-F_1^AF_2^A+(F_1^V+2F_2^V)F_1^A\right]
  \delta^6+O(\delta^7)\Bigg\}.
\end{eqnarray}
It is important to realize that we define the two components of the
polarization of the electron in the $q$ frame and {\it not} in the $n$ frame.
The authors of Ref.~\cite{Hagiwara:1989zt} have shown how to convert the two
polarization components from one frame to the other.

For completeness we present the numerator expression for the normal
polarization of the electron which is given by
\begin{equation}
\label{normpol}
\frac{d\Gamma(P_e^n)}{dq^2} = -\frac{3\pi}{2}
  \frac{\Gamma_0(q^2-m_e^2)^2p}{M_n^7q^2}
  \sqrt{\frac{\delta_e}{2}}{\cal H}_{ISL_+}.
\end{equation}
The normal polarization $P_e^n$ is a $T$-odd observable and is thus
contributed to by the imaginary part of the bilinear helicity forms as shown
in Eq.~(\ref{normpol}). The corresponding triple momentum product can be seen
to be $(\hat p_e \times \hat p_p)\cdot \hat s_e$.

\section{Electron energy distributions\label{sec7}}
One can turn the differential $\cos\theta$ distributions used in the cascade
approach into differential $E_e$ distributions in the direct decay approach
employing the relations
\begin{equation}
\label{transtheta1}
\cos\theta = \frac{2q^2 E_e-q_0(q^2+m_e^2)}{p(q^2-m_e^2)}\,, \qquad \qquad
d\cos\theta=dE_e\,\frac{2q^2}{p(q^2-m_e^2)}
\end{equation}
($q_0=(M_n^2-M_p^2+q^2)/2M_n$), where $E_e$ is the energy of the electron in
the $n$ frame ($m_e \le E_e \le (M_n^2-M_p^2+m_e^2)/2M_n$). The first relation
of Eq.~(\ref{transtheta1}) can be obtained by evaluating the scalar product
$p_n\cdot p_e$ both in the $q$ frame and in the $n$ frame. The relevant
four-vectors in the two frames read
\begin{eqnarray}
    \mbox{\rm $n$ frame}: \qquad  
    && p_n=(M_n;0,0,0)\qquad \qquad \mbox{\rm $q$ frame}:\qquad
    p_n=M_n/\sqrt{q^2}\,(q_0;0,0,p)\nn
    &&p_e=(E_e;p_{e\,x},0,p_{e\,z}) \qquad \qquad \qquad \qquad \ \
    p_e= (\tilde E_e; | \vec{\tilde {p}}_e|\sin\theta,
    - |\,\vec{\tilde{p}}_e|\cos\theta)
\end{eqnarray}
where $\tilde E_e=(q^2+m_e^2)/2\sqrt{q^2}$ and
$|\,\vec{\tilde{p}}_e|=(q^2-m_e^2)/2\sqrt{q^2}$ are the energy and magnitude
of the three-momentum of the electron in the $q$ frame. One then arrives
at Eq.~(\ref{transtheta1}).

Next we consider the azimuthally integrated form of Eq.~(\ref{angdist4}) and
effect the change of variables given in Eq.~(\ref{transtheta}). For the
$(q^2,E_e)$ distribution one obtains
\begin{eqnarray}
\label{q2Eedist}
\frac{d\Gamma}{dq^2dE_ed\cos\theta_P}&=&\frac{\Gamma_0}{M_n^7}
\Big(A_0(q^2,m_e)+A_1(q^2,m_e)E_e+A_2(q^2)E_e^2\strut\nonumber\\&&\strut
  +P_n\cos\theta_P\left(A_0^P(q^2,m_e)+A_1^P(q^2,m_e)E_e+A_2^P(q^2)E_e^2
  \right)\Big)\,.
\end{eqnarray}

Quite remarkably, the coefficients of the quadratic energy dependence
$A_2(q^2)$ and $A_2^P(q^2)$ depend only on $q^2$ and not on the mass $m_e$ of
the lepton~\cite{Penalva:2019rgt}. For these two coefficients one finds
\begin{eqnarray}
\label{A2}
A_2(q^2)&=& \,\frac{3q^2}{2p^2}({\cal H}_U-2{\cal H}_L)\,, \nn
A_2^P(q^2)&=&-\,\frac{3q^2}{2p^2}({\cal H}_F+2{\cal H}_{L_-})\,.
\end{eqnarray}
An explicit calculation shows that $({\cal H}_F+2{\cal H}_{L_-})\sim p^2$
which will compensate the $p^2$ factor in the denominator of the polarized
term $A_2^P(q^2)$ in Eq.~(\ref{A2}). The unpolarized term $A_2(q^2)$
proportional to  $({\cal H}_U-2{\cal H}_L)\sim p^3$ even vanishes at the zero
recoil point $q^2=(M_n-M_p)^2$ and $E_e=((M_n-M_p)^2+m_e^2)/(2(M_n-M_p))$.

The remaining coefficients in (\ref{q2Eedist}) are given by
\begin{eqnarray}
A_0(q^2,m_e)&=&\frac3{8q^2p^2}\Big(q^2(q^2+m_e^2)(2p^2+q^2+m_e^2)
  \left({\cal H}_U-2{\cal H}_L\right)\strut\nn&&\strut
  +2p^2(q^2-m_e^2)\left((2q^2+m_e^2){\cal H}_L+m_e^2{\cal H}_S\right)
  +2q_0p(q^2+m_e^2)\left(q^2{\cal H}_F+2m_e^2{\cal H}_{SL+}\right)\Big)\,,\nn
A_1(q^2,m_e)&=&\frac{-3}{2p^2}\Big(q_0(q^2+m_e^2)
  \left({\cal H}_U-2{\cal H}_L\right)
  +p\left(q^2{\cal H}_F+2m_e^2{\cal H}_{SL+}\right)\Big)\,,\nn
A_0^P(q^2,m_e)&=&\frac{-3}{8q^2p^2}\Big(q^2(q^2+m_e^2)(2p^2+q^2+m_e^2)
  \left({\cal H}_F+2{\cal H}_{L-}\right)\strut\nn&&\strut
  -2p^2(q^2-m_e^2)\left((2q^2+m_e^2){\cal H}_{L-}+m_e^2{\cal H}_{S-}\right)
  +2q_0p(q^2+m_e^2)\left(q^2{\cal H}_U-2m_e^2{\cal H}_{SL-}\right)\Big)\,,\nn
A_1^P(q^2,m_e)&=&\frac{3}{2p^2}\Big(q_0(q^2+m_e^2)
  \left({\cal H}_F+2{\cal H}_{L-}\right)
  +p\left(q^2{\cal H}_U-2m_e^2{\cal H}_{SL-}\right)\Big)\,.\qquad
\end{eqnarray}

The two-fold distribution~(\ref{q2Eedist}) can be further integrated over
$q^2$ or $E_e$ where the respective limits of integration can be derived
from Eq.~(\ref{transtheta}) by setting $\cos\theta=\pm1$. They read
\begin{equation}
\label{limit1}
E_e^\pm = \frac{1}{2q^2}\Big(q_0(q^2+m_e^2) \pm p(q^2-m_e^2)\Big)
\end{equation}
and 
\begin{eqnarray}
\label{limit2}
q^2_{\pm} &=&\frac{M_n(M_n^2-M_p^2+m_e^2-2M_nE_e)
\Big( E_e \pm \sqrt{E_e^2-m_e^2}\Big)}{M_n^2+m_e^2-2M_nE_e} \nn
&=&\frac{2M_n^2}{2M_n(E_e^{\rm max}-E_e)+M_p^2}\left((E_e^{\rm max}-E_e)
\bigg(E_e \pm \sqrt{E_e^2-m_e^2}\,\,\bigg)
+ \frac{m_e^2M_p^2}{2M_n^2} \right)\,,
\end{eqnarray}
where $E_e^{\rm max}=(M_n^2-M_p^2+m_e^2)/2M_n$. Integrating the
distribution~(\ref{q2Eedist}) over $E_e$ in the limits~(\ref{limit1}) one
obtains the one-fold $q^2$ distribution discussed in Sec.~\ref{sec3}. On the
other hand, integrating (\ref{q2Eedist}) over $q^2$ in the
limits~(\ref{limit2}) one obtains the one-fold $E_e$ distribution discussed in
Refs.~\cite{Bender:1968zz,Wilkinson:1982hu} for the unpolarized case.

  \section{Numerical results\label{sec8}}
\begin{table}
\begin{center}
\bgroup
\setlength\tabcolsep{0.2cm}
\renewcommand{\arraystretch}{1.8}
\begin{tabular}{|c|p{6.0cm}|c|c|r|c|}\hline
Observable & LO result & full result & LO value & $\delta_i$(NLO)
& $\langle A_i\rangle '/\langle A_i\rangle$\,(LO)\ \\ \hline
$\langle A_{\rm FB} \rangle$ 
& $\left (|F_1^A|^2 + |F_1^V|^2 \right ) r_{1}(x) /R(x)$ 
& $-0.1440$ & $-0.1448$ & $-5.35\permille$ & $-0.332$ \\ \hline
$\langle A_{\rm conv} \rangle$ 
& $\left ((F_1^A)^2 + (F_1^V)^2 \right ) r_{2}(x)/R(x)$
& $-0.04771$ & $-0.04771$ & $0\permille$ & $-0.332$ \\ \hline
$\langle A_{\rm PFB} \rangle/P_n$
& $F_1^V F_1^A \, r_{3}(x)/R(x)$
& $+0.2382$ & $+0.2388$ & $-2.47\permille$ & $-0.518$ \\ \hline
$\langle A_{\rm DFB} \rangle/P_n$
& $\left( |F_1^A|^2 \,r_{4}(x)  + F_1^V F_1^A \,r_{4'}(x)\right)/R(x)$
& $-0.2159$ & $-0.2159$ & $0\permille$ & $-0.145$ \\ \hline
$\langle A_{\rm Pconv} \rangle/P_n$ 
& $F_1^V F_1^A \, r_{5}(x)/R(x)$
& $-0.03686$ & $-0.03679$ & $+1.88\permille$ & $-0.518$ \\ \hline
$\langle A_{\chi_1} \rangle/P_n$
& $ \pi\left ( |F_1^A|^2 \, r_{6}(x) + F_1^V F_1^A \, r_{6}'(x) \right )/R(x)$
& $+0.3351$ & $+0.3348$ & $+0.85\permille$ & $+0.129$ \\ \hline
$\langle {A}_{\chi_2}\rangle/P_n$
& $F_1^V F_1^A \, r_{7}(x)/R(x)$
& $+0.03705$ & $+0.03693$ & $+3.24\permille$ & $-0.518$ \\ \hline
$\langle P_e^\ell \rangle$
& $\left(|F_1^A|^2 r_{8}(x)+|F_1^V|^2r_{8}'(x)\right)/R(x)$
& $-0.4964$ & $-0.4964$ & $0\permille$ & $+0.337$ \\\hline
$\langle P_e^t \rangle$
& $\pi$ $ (|F_1^A|^2 + |F_1^V|^2 )r_{9}(x)/R(x)$
& $+0.3937$ & $+0.3931$ & $+1.63\permille$ & $-0.332$ \\ \hline 
\end{tabular}
\caption{\label{as}Asymmetries in neutron $\beta$ decay. First column:
Asymmetry; Second column: Analytical expression for LO result
$\Gamma_i^{(5)}/\Gamma^{(5)}$; Third column: Full result; Fourth column:
numerical value of the LO result; Fifth column: numerical value for the
relative NLO correction; Sixth column: error propagation factor.}
\egroup
\end{center}
\end{table}

In Table~\ref{as} we list our analytical and numerical results for the nine
average asymmetries calculated in this paper where we include the two
polarization components of the electron in the list of the asymmetries
since the polarization components are frequently referred to as polarization
asymmetries in the literature. In order to simplify the discussion we set
$P_n=1$ for the five polarization observables, i.e.\ we set
$\langle A_i \rangle /P_n=\langle A_i \rangle$ for the five polarization
observables $\langle A_{\rm PFB} \rangle$ to $\langle A_{\chi_2} \rangle$.

Column~2 contains our analytical LO results for the average asymmetries
$\langle A_i \rangle = P_i\, \langle {\cal O}_i\rangle
=P_i\,\Gamma_i^{(5)}/\Gamma^{(5)}$ where the factors $P_i$ are the same as in
Eqs.~(\ref{FB1}) and~(\ref{FB2}), $P_\ell=P_t=1$. We have cancelled some
numerical factors in the ratio expressions which are now normalized to the
rate factor
\begin{equation}
R(x)=\left(3 (F_1^A)^2 + (F_1^V)^2\right)r(x),
\end{equation}
where $r(x)$ is listed in Eq.~(\ref{rofx}). The LO contributions in column~2
are written in terms of a number of $x$-dependent functions
$r_{1}(x)$ to $r_{9}(x)$ which are defined by
\begin{eqnarray}
r_1(x) &=& \frac{15}{8}x^2\left[(1-x^2)(5+x^2)+4(1+2x^2)\ln x\right]
  \ =\ -0.1531, \nn
r_2(x) &=& -\frac{3}{64}\left[\sqrt{1-x^2}\left(8+194x^2+113x^4\right)
  +15x^2\left(8+12x^2+x^4\right)L(x)\right]
  \ =\ -0.05042, \nn
r_3(x) &=& \frac54\left[\left(1-x^2\right)\left(1-5x^2-2x^4\right)
-12x^4\ln x\right]\ =\ 0.5196, \nn
r_4(x)&=& -\frac12r(x)\ =\ -0.2363, \nn
r_4'(x) &=& \frac5{16}x^2\left[\sqrt{1-x^2}\left(46+29x^2\right)
  +3\left(8+16x^2+x^4\right)L(x)\right]
  \ =\ -0.1691, \nn
r_5(x) &=& -\frac{15}{32}\left[\left(1-x^2\right)\left(1+10x^2+x^4\right)
  +12x^2(1+x^2)\ln x\right]\ =\ -0.08004, \nn
r_6(x) &=& \frac{15}{64}\left[x\sqrt{1-x^2}\left(1+14x^2\right)
  +\left(1-8x^2-8x^4\right)\arccos x\right]\ =\ 0.1498, \nn    
r_6'(x) &=& -\frac5{16}x^2\left[x\sqrt{1-x^2}\left(13+2x^2\right)
  -3\left(1+4x^2\right)\arccos x\right]\ =\ 0.04116, \nn
r_7(x) &=& \frac13(1-x)^5(1+5x)\ =\ 0.08033, \nn
r_8(x) &=& -\frac16\left[\sqrt{1-x^2}\left(18+139x^2+8x^4\right)
  +15x^2\left(8+3x^2\right)L(x)\right]\ =\ -0.8855, \nn
r_8'(x) &=& -\frac16\left[\sqrt{1-x^2}\left(6+193x^2+56x^4\right)
  +15x^2\left(8+9x^2\right)L(x)\right]\ =\ 0.05966, \nn
r_9(x) &=& \frac52x(1-x)^4\ =\ 0.1322.
\end{eqnarray}
In column~3 we list numerical values for the full results using the form
factor values specified in Eqs.~(\ref{marshak}) and~(\ref{F1A}). The full
values are calculated prior to the expansion in $\delta$, not taking into
account the $q^2$ dependence of the form factors. This $q^2$ dependence of
the form factors effects the result far below the precision given in
Tab.~\ref{as}. The predicted values for the average asymmetries range from
$\langle {A}_{\chi_2}\rangle=0.03705 $ to $\langle {P}_e^\ell\rangle=-0.4964$.
The small value of $\langle {A}_{\chi_2}\rangle=0.03705$ results in part from
the smallness of the sector projection factor $P_{{A}_{\chi_2}}=1/3$. In
column~4 we write down the LO numerical values of the analytical LO results
in column~3. The LO values can be seen to be quite close to the full results.

In order to check on the magnitude of the NLO corrections we list the
numerical values for the relative NLO corrections $\delta_i$(NLO) im column~4.
According to Eq.~(\ref{recexp}) the NLO corrections are given by
$\delta_i({\rm NLO})=\Gamma_i^{(6)}/\Gamma_i^{(5)}\,\delta$. The analytical
expressions for the NLO corrections can be found in
Eqs.~(\ref{FB}--\ref{chi2}) and Eqs.~(\ref{epoll2}--\ref{epolt}) and have been
evaluated with the form factor values listed in Eqs.~(\ref{marshak})
and~(\ref{F1A}). The NLO corrections to the LO results listed in column~5 are
generally quite small or even zero. The largest NLO correction occurs for the
forward--backward asymmetry $\langle F_{\rm FB} \rangle$ with
$\delta_{\rm FB}({\rm NLO})=-5.35 \permille$. The NLO corrections move the LO
values very close to the full result in column~3 which shows that one can
safely truncate the recoil expansion at NLO. 

Our results on the polarization observable $\langle A_{\rm PFB} \rangle$ can
be directly compared to the experiment since the average asymmetry
$\langle A_{\rm PFB} \rangle$ is identical to the so-called proton asymmetry
parameter $C$ in the conventional approach. The parameter $C$ has been
measured by the PERKEO~II collaboration with the result
$C=-0.2377(26)$~\cite{Schumann:2007hz}. This value is quite compatible with
our full result $\langle A_{\rm PFB} \rangle=C=-0.2382$. The relative NLO
correction $\delta_i\, ({\rm NLO})=-2.47 \permille$ shifts the LO result
$\langle A_{\rm PFB} \rangle ({\rm LO})=0.2328$ close to the central
experimental value.

It is interesting to know how an error in the value of the axial form factor
propagates to the average asymmetries. This bears on the question on how
accurately can one determine the value of the axial form factor from a
measurement of the average asymmetries discussed in this paper. We discuss
this issue using the usual ratio $\lambda=F_1^A/F_1^V$. Expanding the
asymmetry around the central value $\lambda=1.2724$ from
Ref.~\cite{Tanabashi:2018oca}, one has
\begin{equation}
\label{expansion}
  \langle A_i \rangle (\lambda +\Delta\lambda)=
  \langle A_i \rangle (\lambda)\Big(1 + \frac{\langle A_i \rangle'(\lambda)}
          {\langle A_i \rangle(\lambda)}\cdot\Delta\lambda\Big).
\end{equation}
The experimental value of $\Delta\lambda=0.0023$~\cite{Tanabashi:2018oca} is
small enough that we can terminate the Taylor expansion after the linear term.
The relative error of the average asymmetry $\langle A_i \rangle$ is given by
$\delta\langle A_i \rangle= \langle A_i \rangle'(\lambda)/
\langle A_i \rangle(\lambda)\cdot\Delta\lambda$. The ratio
\begin{equation}
\label{errorp}
\frac {\delta\langle A_i \rangle}{\Delta\lambda}
  = \frac{\langle A_i \rangle'(\lambda)}{\langle A_i \rangle(\lambda)}
\end{equation}
provides a measure of the error propagation from the absolute error of
$\lambda$ to the relative error of the asymmetry $\langle A_i\rangle(\lambda)$.
One wants the error propagation factor to be as large as possible. Of course,
one can turn this argument around. The error propagation from the the relative
error of the asymmetry $\langle A_i\rangle(\lambda)$ to the absolute error of
$\lambda$ is given by the inverse of Eq.~(\ref{errorp}). A good asymmetry
measurement is characterized by a small value of the inverse of
Eq.~(\ref{errorp}). In column~6 we have listed the LO values of the ratio
$\langle A_i \rangle'(\lambda)/\langle A_i \rangle(\lambda)$ where we take the
central PDG value $\lambda =1.2724$. The error propagation factor ranges from
$0.129$ for $\langle A_{\chi_1} \rangle$ to $0.518$ for
$\langle A_{\rm PFB}\rangle,\, \langle A_{\rm Pconv}\rangle$ and
$\langle A_{\chi_2}\rangle$. The latter three asymmetries are thus the best
candidates for an accurate measurement of the axial form factor $F_1^A$. These
three asymmetries would have to be measured with an error less than
$1.19\permille$ to reduce the present PDG error on $\lambda$ given by
$2.3\permille$.

It is interesting to compare the error propagation of $\lambda$ into the total
rate where one has
$\Gamma'(\lambda)/\Gamma(\lambda) = 6\lambda/((1+3\lambda^2)=1.303$. As
concerns the error propagation, the rate measurement is $2.5$ times better
than the best asymmetry measurement. However, the extraction of $\lambda$ from
the rate measurement requires additional input in the form of the value of
$V_{ud}$ and the size of the radiative corrections~\cite{Czarnecki:2019mwq}.
In contrast to this the asymmetry measurements are independent of the value of
$V_{ud}$. Furthermore, the bulk of the radiative corrections can be expected
to cancel out in the asymmetry ratios.
  
\section{Summary and conclusion\label{sec9}}
We have presented the results of a detailed analysis of unpolarized and
polarized neutron $\beta$ decays in the helicity framework. We have derived
exact relativistic formulas for the $q^2$ distribution of the total rate and
the partial correlation rates without employing any recoil approximations. The
$q^2$ integration of the differential rates was done analytically, and the
results were checked by numerical integration. After the $q^2$ integration we
performed an expansion in the small recoil parameter
$\delta=(M_n-M_p)/(M_n+M_p)= 0.689\cdot 10^{-3}$, the series of which has very
rapid convergence properties. Doing the recoil expansion after the integration
spares one from having to guess to which order a given term will contribute
to the final result before doing the final integration. We found that the NLO
term in the recoil expansion vanish for three of the four p.c.\ observables
analyzed in this paper. These are
$\langle A_{\rm conv} \rangle$, $\langle A_{\rm DFB} \rangle$ and the average
value of the longitudinal polarization $P_e^\ell$ of the electron.

At the LO of the recoil expansion one has contributions only from the form
factors $F_1^V$ and $F_1^A$. This opens the opportunity for further
measurements of the form factor $F_1^A$ from other observables on top of the
usual determination of $F_1^A$ from the rate measurement (see the discussion
in Ref.~\cite{Czarnecki:2018okw,Czarnecki:2019mwq}). Particularly well suited
for such a  measurement of $F_1^A$ would be the three observables
$\langle A_{\rm PFB}\rangle$, $\langle A_{\rm Pconv}\rangle$ and
$\langle A_{\chi_2}\rangle$ which are the most sensitive asymmetries for a
determination of $F_1^A$. We find that there is no possibility to measure the
value of the form factor $F_3^A$ nor the slope of the form factors
$F_1^V(q^2)$ or $F_1^V(q^2)$ close to origin since both contribute only to
higher orders in the recoil expansion.

Some of our results are directly applicable to results derived in the
conventional three-body decay analysis done in the $n$ frame. Very obviously,
this holds true for the total rate and the spin--momentum correlation between
the spin of the neutron and the momentum of the proton conventionally called
the spin--proton correlation parameter $C$. The spin--electron and
spin--neutrino correlations defined in the conventional approach are not
directly related to the corresponding correlations in the helicity approach.
As concerns azimuthal correlations one can choose the momentum of the proton
to define the $z$ axis in the direct decay approach (system~2 in
Ref.~\cite{Korner:1998nc}). For this choice the azimuthal correlations in the
two approaches are simply related. As shown in Sec.~\ref{sec3}, the $T$-odd
triple correlation parameter $D$ of the conventional approach is proportional
to $(-1/2\, {\cal H}_{ILT_-}+ \delta_e{\cal H}_{IST_+})$ in the helicity
approach. The same holds true for the $T$-odd normal polarization $P_e^n$ of
the electron discussed in Sec.~\ref{sec7}.

As discussed in Sec.~\ref{sec7}, one can turn the differential $\cos\theta$
distribution used in the helicity approach into a differential electron energy
distribution in the conventional direct decay approach employing the
relation~(\ref{transtheta}),
\begin{equation}
\cos\theta = \frac{2q^2 E_e-q_0(q^2+m_e^2)}{p(q^2-m_e^2)}\,,
\end{equation}
where $E_e$ is the energy of the electron in the neutron rest frame.

Other results of the conventional three-body decay analysis such as  opening
angle distributions between pairs of the three final state particles
$(p,\,e^-,\,\bar \nu)$ in the neutron rest frame are not part of the helicity
analysis. These distributions can be obtained by applying the appropiate
boosts to the helicity distributions either analytically or by Monte Carlo
event generation methods as has been done in the analysis of polarized hyperon
decays $\Xi^0 \to \Sigma^+ + \ell^- +\bar \nu_\ell\,\, (\ell^-=e^-,\,\mu^-)$
in Ref.~\cite{Kadeer:2005aq}.

One of the advantages of using normalized angular observables is that they do
not depend on the value of $|V_{ud}|$ which is welcome even if the relative
error on $|V_{ud}|$ is small
($\sim$ $0.1$\textperthousand~\cite{Tanabashi:2018oca}). Furthermore, the bulk
of the radiative corrections can be expected to cancel when taking ratios of
rates since large parts of the radiative corrections are proportional to the
Born term rates.

In this paper we have restricted our discussion to the helicity analysis of
free neutron $\beta$ decays. There is no obstacle to also apply the helicity
method to nuclear $\beta$ decays.

The results of this paper can also be formulated in terms of an effective
field theory (EFT) approach (see e.g.\ Ref.~\cite{Gonzalez-Alonso:2018omy}).
In addition, New Physics effects (see e.g.\ Ref.~\cite{Cirgiliano:2019nyn})
are easily incorporated into the helicity framework. An EFT helicity approach
to neutron $\beta$ decay including New Physics effects will be the subject of
a sequel to this paper. 

\subsection*{Acknowledgments}
We would like to thank J.~Erler, W.~Heil, D.~McKay, W.~Shepherd for
discussions and encouragement. B.M.\ has been supported by the European Union
through the European Regional Development Fund -- the Competitiveness and
Cohesion Operational Programme (KK.01.1.1.06). B.M.\ would like to acknowledge
the support of the Alexander von Humboldt foundation as well as the
hospitality of the theory group THEP at the Institute of Physics at the
Johannes Gutenberg University. The research of S.G.\ was supported by the
European Regional Development Fund under Grant No.~TK133. S.G.\ also
acknowledges support from the PRISMA and PRISMA$^+$ (project No.~2118 and ID
39083149) Clusters of Excellence at the University of Mainz and the
hospitality of the Institute for Theoretical Physics at the University of
Mainz.

\end{document}